\documentclass[twocolumn]{aastex631}
\usepackage{amsmath}
\usepackage{amssymb}
\usepackage{mathrsfs}
\usepackage{xcolor}
\usepackage{hyperref}
\usepackage{physics}
\usepackage{simplewick}
\usepackage{epsfig}
\usepackage{array}
\usepackage{booktabs}
\usepackage{subfigure}
\usepackage{gensymb}
\usepackage{tabularx}
\usepackage{makecell}

\colorlet{DeepBlue}{blue!80!black} 
\newcommand{\B}[1]{\textcolor{DeepBlue}{#1}}

\begin{document}

\title{Simulations of interaction between outflow and surrounding broken power-law circumnuclear medium: implications for different radio light curves of TDEs 
}

\author[0009-0003-0516-5074]{Xiangli Lei}
\affiliation{Department of Astronomy, School of Physics, Huazhong University of Science and Technology, Luoyu Road 1037, Wuhan, China}

\author[0000-0003-4773-4987]{Qingwen Wu$^*$}
\email{* Corresponding author: qwwu@hust.edu.cn} 
\affiliation{Department of Astronomy, School of Physics, Huazhong University of Science and Technology, Luoyu Road 1037, Wuhan, China}

\author[0009-0005-9790-1263]{Chang Zhou}
\affiliation{Department of Astronomy, School of Physics, Huazhong University of Science and Technology, Luoyu Road 1037, Wuhan, China}

\author[0000-0003-3440-1526]{Wei-Hua Lei}
\affiliation{Department of Astronomy, School of Physics, Huazhong University of Science and Technology, Luoyu Road 1037, Wuhan, China}

\author[0000-0002-7329-9344]{Ya-Ping Li}
\affiliation{Shanghai Astronomical Observatory, Chinese Academy of Sciences, Shanghai 200030, People’s Republic of China}

\author[0000-0002-2581-8154]{Jiancheng Wu}
\affiliation{Department of Astronomy, School of Physics, Huazhong University of Science and Technology, Luoyu Road 1037, Wuhan, China}

\author[0009-0002-3470-5052]{Weibo Yang}
\affiliation{Department of Astronomy, School of Physics, Huazhong University of Science and Technology, Luoyu Road 1037, Wuhan, China}

\begin{abstract}

The complex radio light curves of tidal disruption events (TDEs) challenge our understanding of the properties of both the outflows and the circumnuclear medium (CNM) surrounding supermassive black holes. In this work, we explore outflow-CNM interactions across a broad parameter space using three-dimensional hydrodynamic simulations, adopting a broken power-law CNM density profile with a transition near the Bondi radius. The outflow-CNM interaction inside Bondi radius produces an early radio flare (\(\lesssim 2\) yr) once the emitting region becomes optically thin. A second radio rebrightening can appear a few years later if the outflow decelerates beyond Bondi radius. We also find that either a very dense inner CNM, which causes rapid deceleration, or a rarefied outer CNM suppresses the late rebrightening that will produces a single early-peaked flare. In contrast, a rarefied CNM inside the Bondi radius suppresses the early flare and yields a single late-peaked event. For the case of very dense CNM at large radii, the interaction will trigger a sharp late-time rise as observed in some TDEs. We further explore the interaction of a relativistic jet with a broken power-law CNM, which can reproduce the characteristic light curves as observed in jetted TDEs without invoking complex jet structure.

\end{abstract}


\section{Introduction} \label{sec:intro}

A tidal disruption event (TDE) occurs when a star enters the tidal radius of a supermassive black hole (SMBH) in a galactic nucleus \citep[e.g.,][]{Rees1988, Evans1989, Gezari2021}, producing a luminous, multi-wavelength flare \citep[e.g.,][]{Alexander2020, van_Velzen2020, Lu2020}. The peak fallback rate of the disrupted stellar debris typically exceeds the Eddington limit \citep[e.g.,][]{Strubbe2009, Strubbe2011, Lodato2011}, leading to the formation of an accretion disk that radiates primarily in soft X-rays \citep[e.g.,][]{Ulmer1999, Roth2016, Stone2016, Auchettl2017, Dai2018, Pasham2018, Curd2019, Saxton2020}. However, most TDEs are discovered through optical surveys, and their optical/UV emission mechanisms remain uncertain. Two leading models have been proposed to explain this emission: (1) reprocessing of X-ray photons by an optically thick envelope, and (2) shock heating from collisions among debris streams \citep[e.g.,][]{Guillochon2013, Jiang2016, Piran2015, Roth2016, Lu2020, van_Velzen2020, Liu2021, Steinberg2024}. Recent wide-field surveys further indicate that TDEs preferentially occur in compact, green-valley galaxies, which suggests that TDE rates are closely regulated by the nuclear environments of their host galaxies \citep[e.g.,][]{Hammerstein2021, Yao2023AT2020vwl, Wang2024}.

The radio survey of TDEs remains incomplete, but rapid follow-up observations have yielded over 50 radio detections \citep[e.g.,][]{Alexander2020,Cendes2024,Zhou2026}. Several jet-dominated events have been identified in recent years, including Sw~J1644+57 \citep[e.g.,][]{Bloom2011,Zauderer2011,Levan2011,Metzger2012}, Sw~J2058+05 \citep[e.g.,][]{Cenko2012}, Sw~J1112$-$82 \citep[e.g.,][]{Brown2015}, and AT~2022cmc \citep[e.g.,][]{Andreoni2022}. The origin of the radio emission of other TDEs is still in hot debate, which typically appears tens to hundreds of days after the optical/X-ray flare \citep[e.g.,][]{Alexander2016,Krolik2016,Pasham2018,Alexander2020,Lu2020,Stein2021}. The explanations for the delayed radio emission include circumnuclear medium (CNM) interaction with an on-axis or off-axis jet \citep[e.g.,][]{Cendes2021,Horesh2021a,Matsumoto2023,Teboul2023,Zhou2024,Sfaradi2024,Sato2024}, non-relativistic winds \citep[e.g.,][]{Alexander2016,Lu2020,Mou2022,Bu2023,Matsumoto2024,Zhuang2025}, unbound debris \citep[e.g.,][]{Krolik2016,Lei2016,Yalinewich2019,Spaulding2022}, or 
delayed jet/outflow formation during the late decaying phase in a low accretion state \citep[e.g.,][]{Horesh2021a,Sfaradi2022,Cendes2022,Matsumoto2023}.

The CNM in galactic centers is likely shaped by the accretion history of the central SMBH. In AGNs, gas and dust are abundant, forming a toroidal obscuring structure in the equatorial plane, where the dense gas regions that emit broad lines and more extended less dense gas that emits narrow lines as described by the AGN unification model \citep[e.g.,][]{Antonucci1993,Urry1995,Netzer2015}. In contrast, low-luminosity AGNs and quiescent galaxies often lack such cold components based on absence of the broad emission lines and mid-infrared signatures, where, however, hot gas can be directly resolved on parsec scales via high-resolution X-ray imaging \citep[e.g.,][]{Wang2013,Russell2015,Wong2014}. Therefore, the CNM evolves significantly over the evolutionary history of the host galaxy. However, owing to limited angular resolution, detailed CNM density profiles remain poorly constrained for most galaxies. The CNM density profile normally differs inside and outside the Bondi radius, where the gravitational potential is dominated by the SMBH and by the host bulge, respectively \citep[e.g.,][]{Bondi1952,Quataert2004,Coughlin2014,Generozov2017,Gillessen2019}. For example, the gas density often follows a power-law profile inside the Bondi radius, while it flattens near the Bondi radius and eventually follows a power-law decline beyond kpc scales (e.g., M87; \citealt{Russell2015}). TDEs provide an opportunity to probe the CNM distribution by fitting radio light curves and spectral energy distributions produced by outflow-CNM interaction \citep[e.g.,][]{Mou2022,Lei2024,Matsumoto2024,Zhuang2025}.

With the growing sample of radio TDEs, the observed radio light curves have revealed a remarkable diversity that challenges simple analytical models. For instance, some sources exhibit radio delays of several years after the optical peak \citep[e.g., PS16dtm;][]{Cendes2024}, others show extremely steep flux rises exceeding $t^3$ \citep[e.g., AT2018hyz;][]{Cendes2022}, and some events display double-peaked radio light curves \citep[e.g., ASASSN-15oi;][]{Horesh2021a}. Although recent analytical work has attempted to interpret some of these complex features \citep[e.g.,][]{Teboul2023,Matsumoto2024,Sato2024,Cendes2025AT2018hyz}, a unified framework remains lacking. Moreover, the possible hydrodynamic instabilities and the evolving geometry of the outflow can further influence the actual radio emission, motivating a multidimensional hydrodynamic treatment beyond one-zone analytical frameworks. In this work, we investigate the resulting radio emission using three-dimensional hydrodynamic simulations in which we vary the outflow properties and the CNM density profiles inside and outside the Bondi radius. Our models can roughly reproduce many key features of observed TDE radio light curves, including single early or late peaks, double-peaked structures, and sharply rising late-time flares. Section~\ref{sec:setup} outlines the numerical setup; Section~\ref{sec:Result} presents the simulation results; Section~\ref{sec:discussion} discusses the correspondence between simulated and observed light curves; and Section~\ref{sec:summary} summarizes the main conclusions.

\section{Numerical Setup} \label{sec:setup}

We model the outflow as a uniform, axisymmetric outflow launched from the inner boundary and simulate its interaction with the CNM. When a star is disrupted and accreted by an SMBH, the resulting super-Eddington accretion flow can launch a fast wind or a relativistic jet \citep[e.g.,][]{Dai2018,Curd2021,Qiao2025,Bu2026}. As these outflows propagate outward, they interact with the surrounding CNM and drive shocks that accelerate nonthermal electrons and produce synchrotron emission. We eject a TDE outflow with a total mass of \(M_{\rm ej}\) from the inner boundary over the interval \(0 < t < t_{\rm ej}\). Within the polar cone \(\theta < \theta_{\rm ej}/2\), the material is ejected with a constant mass-outflow rate \(\dot{M}_{\rm ej}\) at velocity of \(v_{\rm ej}\) during the ejection duration \(t_{\rm ej}=0.1\,\mathrm{yr}\) \citep[e.g.,][]{Li2024,Hayasaki2023}. To avoid numerical instabilities in the simulations, the ejecta temperature is set to \(T_{\rm ej}=10^{8}\,\mathrm{K}\), which roughly do not affect the evolution of shocks and the radio light curves.

During the interaction between the outflow and the CNM, variations in the density profile can significantly modify the resulting radio light curve. To investigate how the CNM structure influences the radio emission, we adopt a simplified broken-power-law density profile with a transition at the Bondi radius,
\begin{equation}
    n(R) =
    \begin{cases}
        n_{\rm B}\,(R/R_{\rm B})^{-k_{\rm i}}, & R \le R_{\rm B},\\
        n_{\rm B}\,(R/R_{\rm B})^{-k_{\rm o}}, & R > R_{\rm B},
    \end{cases}
    \label{eq:nR}
\end{equation}
where \(R_{\rm B}\) is the Bondi radius, \(n_{\rm B}\) is the density at \(R_{\rm B}\), and \(k_{\rm i}\) and \(k_{\rm o}\) describe the power-law slopes of the density profile inside and outside \(R_{\rm B}\), respectively. We fix the Bondi radius at \(R_{\rm B}=1.5\times10^{17}\,\mathrm{cm}\), consistent with observations of Sgr A* \citep{Baganoff2003}. This value corresponds to a black hole mass \(M_{\rm BH}=4\times10^{6}\,M_{\odot}\) and a temperature near the Bondi radius of \(T_{\rm B}\simeq10^{7}\,\mathrm{K}\).

The inner slopes $k_{\rm i}$ in the range from 1.0 to 3.0 are considered, which are constrained from the radio TDE analyses \citep{Alexander2020,Cendes2024} or the directly measured Bondi-scale hot-gas profiles in nearby galactic nuclei \citep[e.g.,][]{Wong2014,Russell2015,Runge2021}. The observed diversities in $k_{\rm i}$ may reflect gas environments and different accretion histories of central supermassive black holes \citep[see review by][]{Alexander2020}. For the outer slope, we consider $k_{\rm o}$ in the range form $-1.0$ to 1.0. The slope $k_{\rm o}<0$ represents the possible local density bump or shell-like enhancement beyond $R_{\rm B}$ where the overdensities can arise from past stellar mass-loss, stalled outflow remnants, or the interface between the hot CNM and cold molecular clouds \citep[as observed in M87, e.g.,][]{Russell2015,Runge2021,Cho2024}.

We employ the publicly available code \texttt{Athena++} \citep[e.g.,][]{Stone2020} to perform three-dimensional hydrodynamic simulations of the interaction between the outflow and the CNM. The simulation domain spans \(r \in [0.01\,R_{\rm B},\,10\,R_{\rm B}]\), \(\theta \in [0,\,\pi/2]\), and \(\phi \in [0,\,2\pi]\), using a logarithmically uniform grid in the radial direction with \(N_r\times N_\theta\times N_\phi = 6144\times64\times256\) cells. Outflow boundary conditions are imposed in the \(r\)-direction, polar boundary conditions in the \(\theta\)-direction, and periodic boundary conditions in the \(\phi\)-direction. The simulation is evolved until the outermost shocked material reaches the outer boundary at \(10\,R_{\rm B}\), which occurs at \(\sim 5\times10^{3}\,\mathrm{days}\) for our fiducial models. The governing equations and the synchrotron-emission calculations are summarized in Appendix~\ref{app:setup}.

\begin{figure*} 
  \centering
  \includegraphics[width=1.0\linewidth]{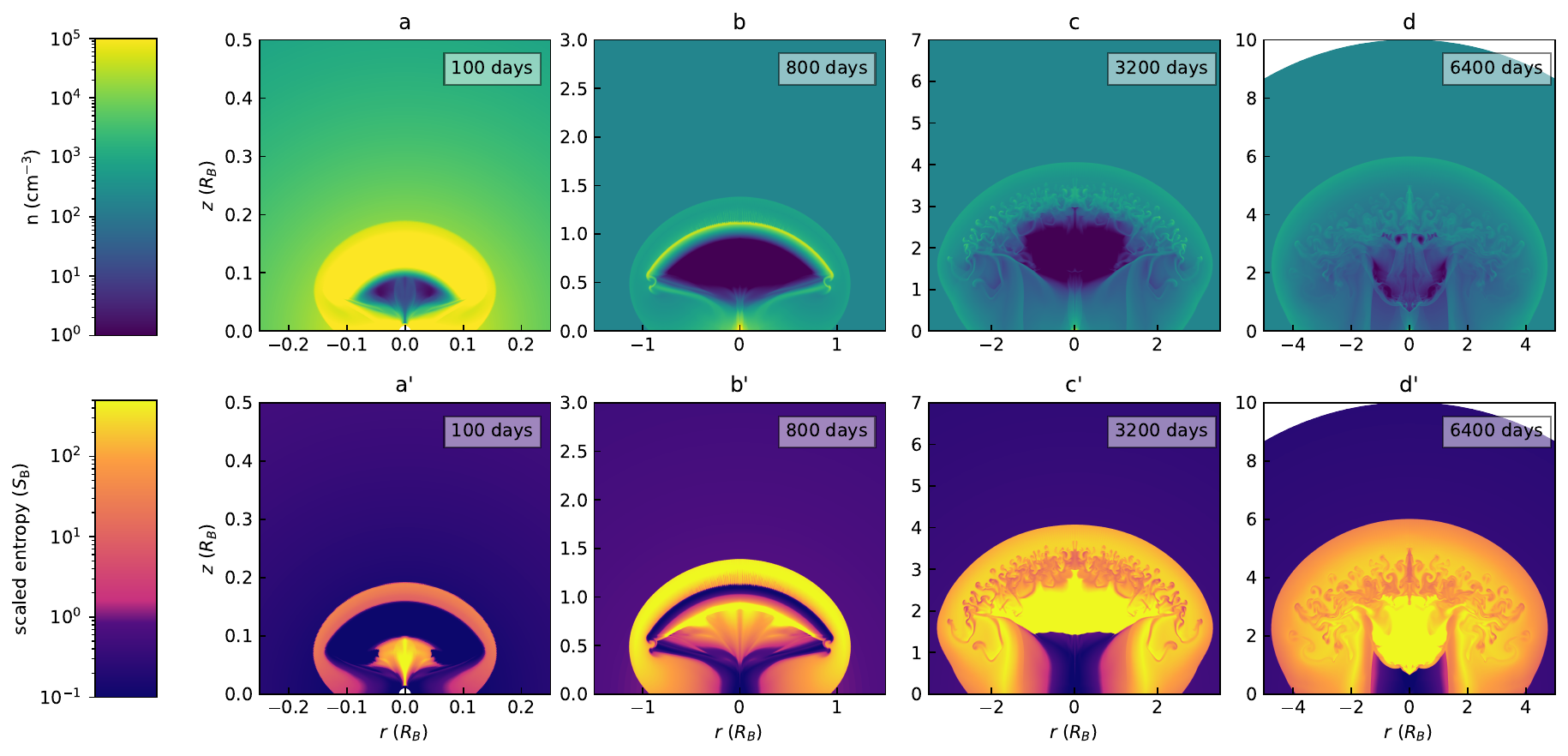}
  \caption{Distributions of the number density (top panel) and scaled entropy (bottom panel) for the wind-CNM interaction with $\theta_{\rm ej}=120^\circ$ for four representative simulation snapshots.}
  \label{fig:w120}
\end{figure*}

\section{Simulation Results} \label{sec:Result}

\subsection{The dynamic evolution} \label{sec:dynamic}

We simulate the interaction between an outflow and a broken-power-law CNM, adopting a fiducial density at the Bondi radius of \(n_{\rm B}=200\ \mathrm{cm^{-3}}\) with \(k_{\rm i}=2.5\) and \(k_{\rm o}=0\) describing the power-law slopes inside and outside \(R_{\rm B}\), respectively. To capture geometric effects, we adopt an opening angle \(\theta_{\rm ej}=120^\circ\) to represent a wide wind with \(M_{\rm ej}=0.02\,M_\odot\) and \(v_{\rm ej}=0.1\,c\).

Figure~\ref{fig:w120} presents snapshots of the number density and the scaled entropy during the wind-CNM interaction. In the early phase (panels a and a'), the wind expands at nearly constant velocity, forming a dense shell with a small fraction of shocked material accumulated at the leading edge. As the forward shock reaches the Bondi radius (panels b and b'), it continues to sweep up ambient material. Once the swept-up mass becomes comparable to the ejecta mass, the flow transitions into the Sedov-Taylor phase \citep[e.g.,][]{Taylor1950,Sedov1959} and begins to decelerate (panels c and c'). When the density of the expanding ejecta drops below that of the CNM, a Rayleigh-Taylor (RT) instability is triggered; this instability disrupts the wind structure and enhances deceleration. The RT-driven disruption and deceleration develop further in panels d and d'.

\subsection{Basic Radio Light Curve} \label{sec:LightCurve}

\begin{figure*} 
  \centering
  \includegraphics[width=1.0\linewidth]{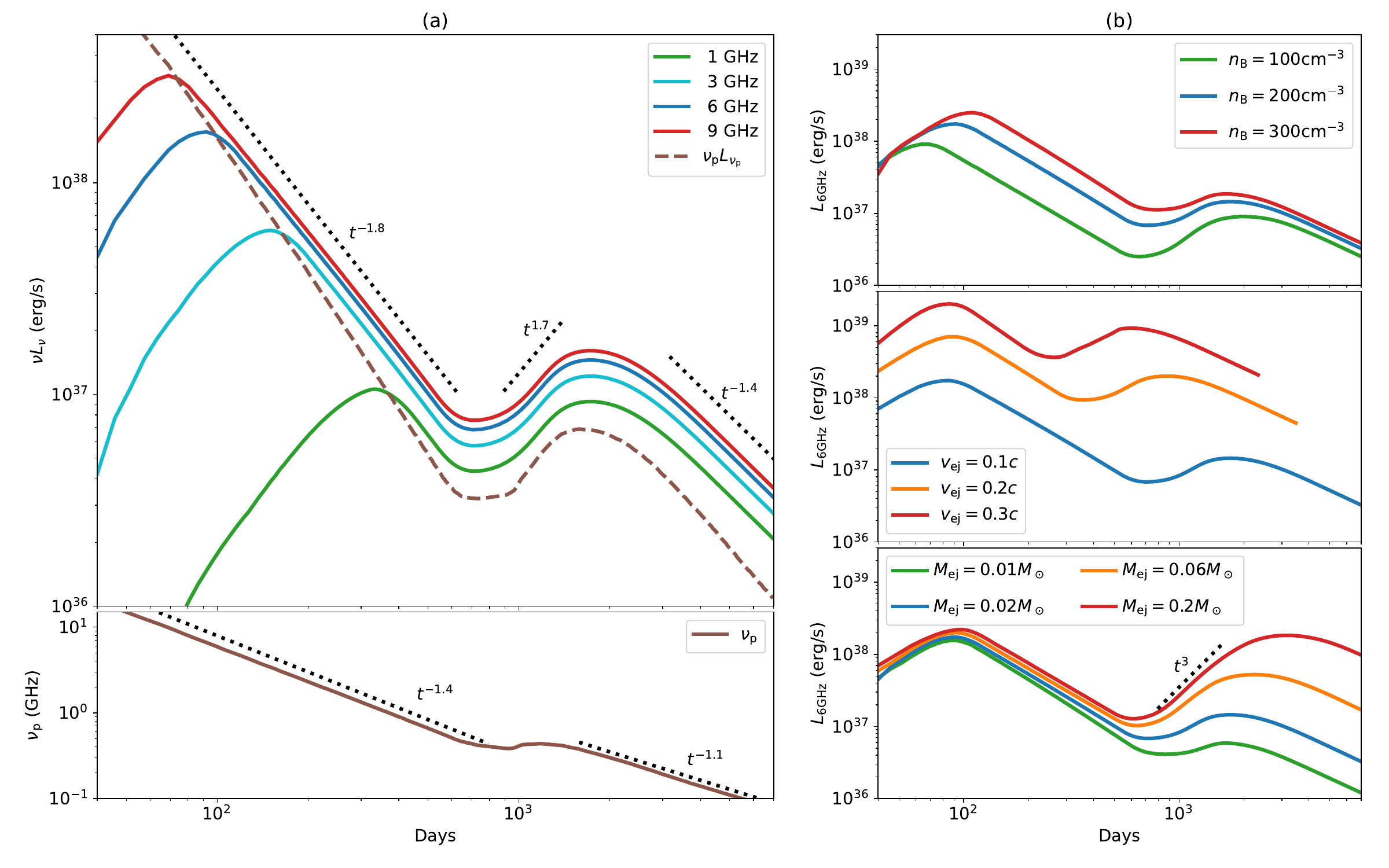}
    \caption{Radio light curves for the wind-CNM interaction. (a) Top-left panel: radio light curves at multiple frequencies (solid lines) for the fiducial parameters \(M_{\rm ej}=0.02\,M_{\odot}\), \(v_{\rm ej}=0.1\,c\), and \(n_{\rm B}=200\,\mathrm{cm^{-3}}\). The dashed line shows the evolution of \(\nu_{\rm p}L_{\nu_{\rm p}}\). The evolution of the peak frequency \(\nu_{\rm p}\) is shown in the bottom-left panel. (b) Right panels: 6-GHz light curves for different model parameters (top: different densities at the Bondi radius; middle: different initial outflow velocities; bottom: different ejecta masses).}
  \label{fig:lc}
\end{figure*}

We simply calculate the synchrotron emission for an observer located along the polar axis, adopting typical parameters \(\epsilon_{\rm e}=0.1\), \(\epsilon_{\rm B}=0.02\), and \(p=2.5\) (see Appendix~\ref{subsec: SynchrotronRadiation} for details). The radio light curves are shown in Figure~\ref{fig:lc}(a). The radio emission exhibits two distinct flares, peaking at \(\sim 10^{2}\) and \(\sim 10^{3}\) days. The first flare is strongly frequency-dependent (higher frequencies peak earlier), whereas the second flare is nearly frequency-independent. The early peak arises from the transition of the emitting region from optically thick to optically thin (see Figure~\ref{fig:lc}(a)), which accounts for its frequency dependence. After the first peak, the luminosity decays approximately as \(\nu L_{\nu}\propto t^{-1.8}\) until the freely expanding ejecta reach the Bondi radius. The light curve shows a break at \(t_{\rm break}\simeq700\) days, coincident with the wind crossing \(R_{\rm B}\). Beyond this time, the radio luminosity rises as \(\nu L_{\nu}\propto t^{1.7}\) because the flatter outer CNM increases the mass-sweeping efficiency. The second peak occurs at \(t_{\rm 2nd}\simeq1750\) days, when the swept-up mass becomes comparable to the ejecta mass. Thereafter, the emission declines as \(\nu L_{\nu}\propto t^{-1.4}\). The second peak is generally dimmer than the first one because less kinetic energy is dissipated at that stage. The spectral peak frequency \(\nu_{\rm p}\), which roughly corresponds to the emission-weighted average of the synchrotron self-absorption frequencies \(\nu_{\rm a}\) of individual shocked cells, evolves as \(\nu_{\rm p}\propto t^{-1.4}\) before the outflow crosses the Bondi radius and transitions to a slower decline of \(\nu_{\rm p}\propto t^{-1.1}\) beyond this radius (bottom-left panel of Figure~\ref{fig:lc}). Notably, \(\nu_{\rm p}\) remains approximately constant for several hundred days after crossing \(R_{\rm B}\). Our simulation results are broadly consistent with the analytical predictions of the spherical thin-shell model \citep{Matsumoto2024}. The simulations provide more detailed results, where the non-negligible kinetic energy dissipation prior to the nominal deceleration radius, multidimensional geometric effects (such as lateral spreading) and RT instabilities modify the dynamical evolution and the resulting radio emission. We also provide analytical estimates for the key timescales, temporal slopes, and peak luminosities in Appendix~\ref{sec:analytic}, which incorporate these corrections and show better agreement with our simulation results.

We further test the model over a broader parameter space using a reduced angular resolution in \(\theta\) and \(\phi\) (from \(256\times64\) to \(128\times32\)), which does not affect our main conclusions. Figure~\ref{fig:lc}(b) presents 6\,GHz light curves for variations in the CNM density \(n_{\rm B}\), ejecta velocity \(v_{\rm ej}\), and ejecta mass \(M_{\rm ej}\). Increasing \(n_{\rm B}\) raises the radio luminosity, delays the first peak, and advances the second peak, reflecting increased mass loading and energy dissipation within \(R_{\rm B}\). Higher \(v_{\rm ej}\) increases the kinetic energy, producing stronger emission and an earlier second flare due to the ejecta reach \(R_{\rm B}\) ealier. Higher \(M_{\rm ej}\) amplifies the post-Bondi interaction, suppresses RT instabilities, and steepens the rise toward \(\nu L_{\nu}\propto t^{3}\) once \(M_{\rm ej}/M_{\rm B}\gtrsim 10\), where \(M_{\rm B}={n_{\rm B}\,m_{\rm p}\,\Omega_{\rm ej}\,R_{\rm B}^{3}}/{(3-k_{\rm i})}\) is the mass swept up by the outflow within the Bondi radius. We find that if the deceleration radius is comparable to or smaller than \(R_{\rm B}\) (i.e., \(M_{\rm ej}/M_{\rm B}\lesssim 1\)), the second peak will be weak or absent owing to the reduced interaction with the outer CNM, yielding a single early peak.

\subsection{Impact of the CNM Density Profile} \label{sec:CNM}

\begin{figure*} 
  \centering
  \includegraphics[width=1.0\linewidth]{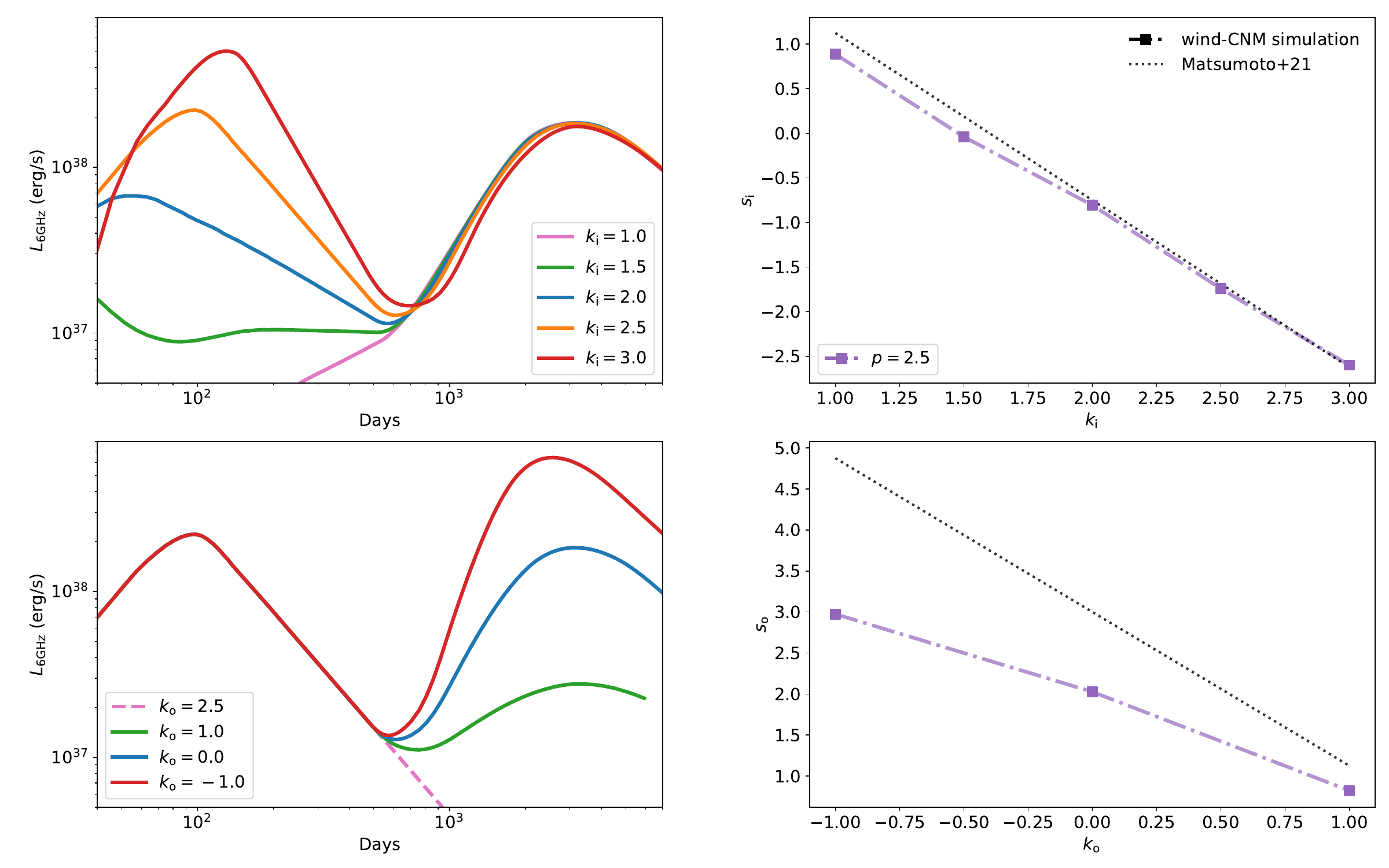}
  \caption{Left panels: Radio light curves for different CNM density profiles inside (top-left panel) and outside the Bondi radius (bottom-left panel), respectively, for \(M_{\rm ej}=0.2\,M_{\odot}\). Right panels: relation between the light-curve slopes and the CNM density slopes; the top-right panel corresponds to the decay phase of the early flare and the bottom-right panel to the rise phase of the late flare. The dotted lines show the analytic solution from \citet{Matsumoto2021}}
  \label{fig:kb}
\end{figure*}

\begin{figure} 
  \centering
  \includegraphics[width=1.0\linewidth]{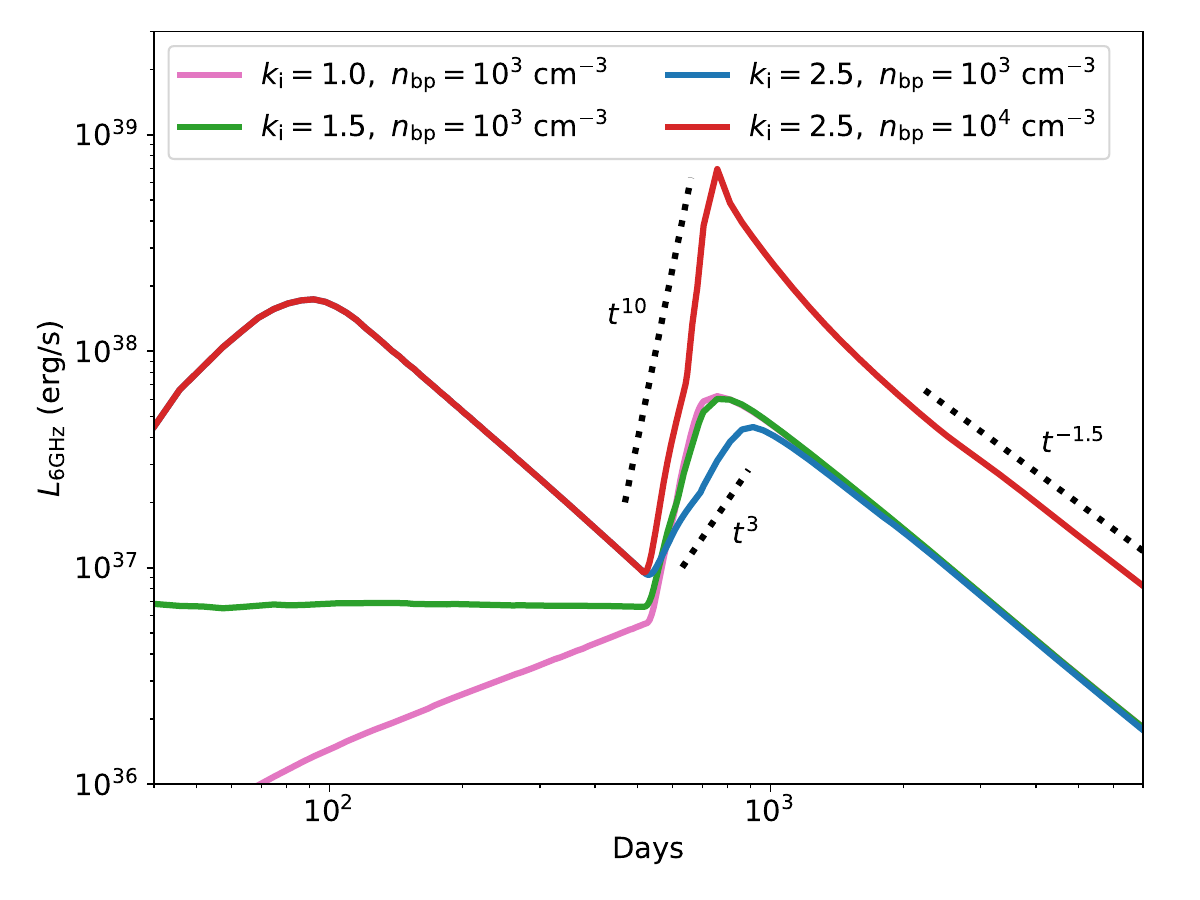}
  \caption{Radio light curves for the wind-CNM interaction (outer dense-gas case) with \(M_{\rm ej}=0.02\,M_{\odot}\), showing the steep rise of the second radio flare.}
  \label{fig:kcold}
\end{figure}

\begin{figure*}
  \centering
  \includegraphics[width=1.0\linewidth]{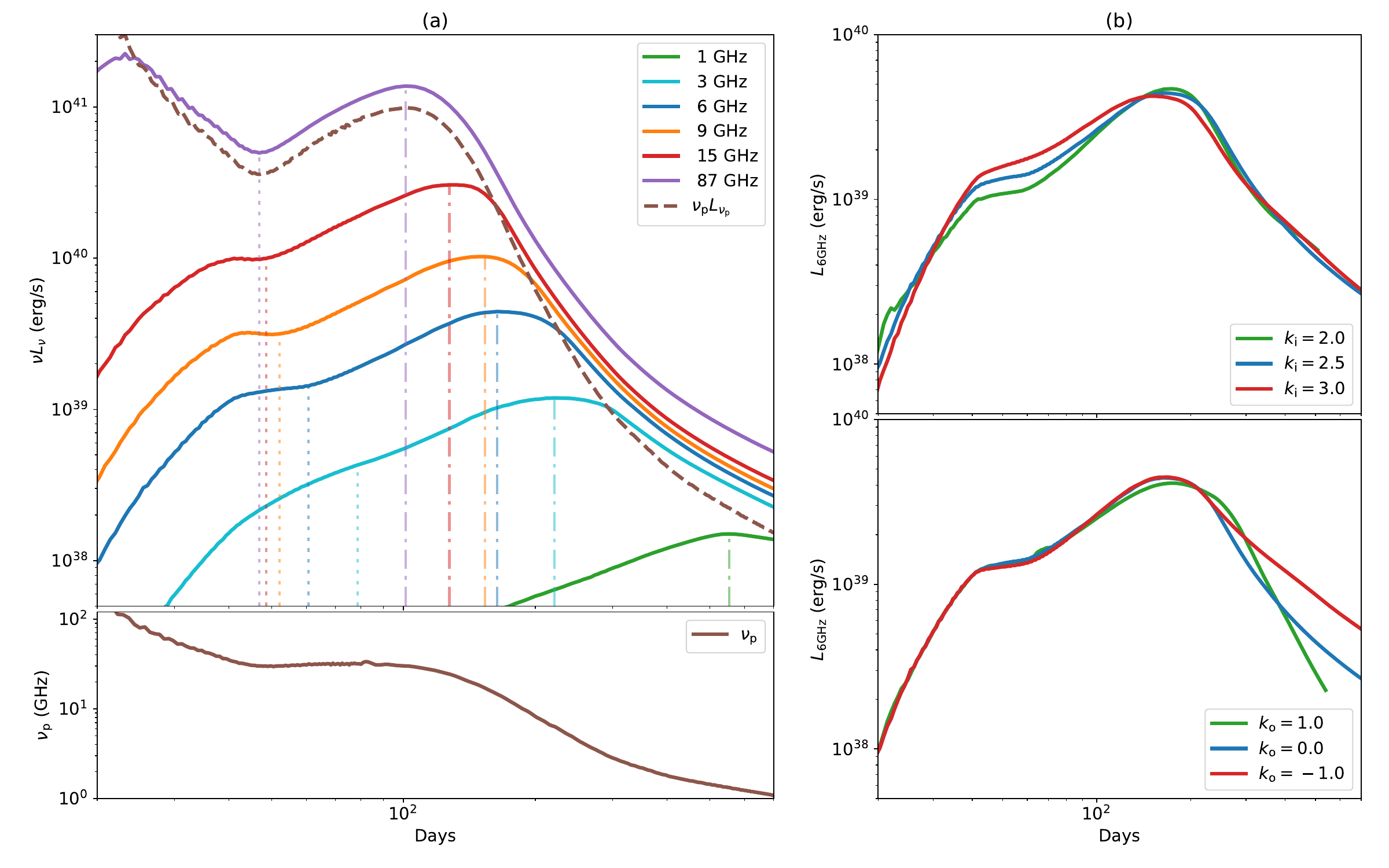}
  \caption{Radio light curves for the jet-CNM interaction. Owing to the high jet velocity, the jet reaches the Bondi radius \(R_{\rm B}\) before the first peak fully develops, preventing a distinct, well-separated double-peaked structure,  where the colored dotted and dot-dashed vertical lines represent \(t_{\rm break}\) and \(t_{\rm 2nd}\) at given radio frequency, respectively. }
  \label{fig:jet_lc}
\end{figure*}

In Section~\ref{sec:LightCurve} we show that outflow-CNM interactions naturally produce double-peaked radio light curves when the deceleration radius greatly exceeds the Bondi radius, corresponding to relatively little mass loaded within the Bondi radius (i.e., \(M_{\rm ej}/M_{\rm B}\gg 1\)). In this regime the light-curve shape is primarily controlled by the radial CNM density profile \citep[see also][for analytical work]{Matsumoto2024}. To quantify this dependence, we define the decay slope inside the Bondi radius by \(\nu L_{\nu}\propto t^{s_{\rm i}}\) and the rise slope of the second flare beyond the Bondi radius by \(\nu L_{\nu}\propto t^{s_{\rm o}}\); both slopes are measured between the 80\% and 20\% flux levels. To better measure the slope of light curve, we adopt \(M_{\rm ej}=0.2\,M_{\odot}\) for the simulations shown in Figure~\ref{fig:kb}, although \(s_{\rm i}\) is largely insensitive to the ejecta mass. A shallower CNM profile inside \(R_{\rm B}\) produces a flatter decline of the first flare because less shocked gas is generated. For \(k_{\rm i}=1.0\), the light curve within \(R_{\rm B}\) can remain in a slowly rising phase. The slope of the second flare correlates strongly with the outer density slope: a rapidly declining CNM (\(k_{\rm o}\gtrsim 1\)) suppresses the second peak, whereas a flat or locally rising profile (\(k_{\rm o}\lesssim 0\)) yields a brighter, more steeply rising late-time flare. The right panels of Figure~\ref{fig:kb} illustrate the relation between the CNM slopes and the measured light-curve slopes. While the simulations agree well with analytic expectations for the first flare, the measured second-flare slopes are systematically shallower than the analytical prediction. This discrepancy is primarily due to the broader spatial expansion of the outflow and the non-negligible kinetic-energy dissipation that occurs prior to reaching the nominal deceleration radius in the simulations. Overall, reduced gas mass inside or outside the Bondi radius weakens the first or second radio peak, respectively.

Apart from the hot CNM, cold clouds may also appear in parsec scales, including components analogous to the AGN broad-line clouds or compact molecular clouds. Observations of nearby nuclei, such as M87, reveal clumpy cold material embedded within the hot atmosphere \citep[e.g.,][]{Russell2018,Simionescu2018}. The density of cold gas \(\sim 10^{4-6}\,\mathrm{cm^{-3}}\), is typically much higher than that of the hot/warm CNM \citep[e.g.,][]{Vollmer2004,Harada2015,Hsieh2021}. To mimic this situation, we consider a simplified scenario in which a density jump dominated by dense gas exists at the Bondi radius \citep[e.g.,][]{Ghizzardi2010,Zinger2018}, followed by a local effective profile with index \(k_{\rm o}= 1.5\) over a limited radial range \citep[e.g.,][]{Vollmer2002,Oka2011,Solanki2023}. Specifically,
\begin{equation}
    n(R) =
    \begin{cases}
        n_{\rm B}\,(R/R_{\rm B})^{-k_{\rm i}}, & R \le R_{\rm B},\\
        n_{\rm bp}\,(R/R_{\rm B})^{-1.5}, & R > R_{\rm B},
    \end{cases}
  \label{eq:nR_jump}
\end{equation}
where \(n_{\rm bp}\) denotes the number density of the bumpy cold gas. The radio light curves from the interaction between the outflow and this bumpy medium are shown in Figure~\ref{fig:kcold}. The density jump at \(R_{\rm B}\) causes rapid accumulation of shocked material as the outflow crosses the Bondi radius, producing a very steep rise, and higher post-jump densities yield correspondingly steeper radio light curves. At fixed \(n_{\rm bp}=10^{3}\,\mathrm{cm^{-3}}\), the case of \(k_{\rm i}=2.5\) gives a second-rise slope only slightly steeper than \(t^{3}\), while the light curves with \(k_{\rm i}=1.0\) and \(k_{\rm i}=1.5\) becomes clearly steeper than \(t^{5}\). Such steep slopes are qualitatively consistent with previous analytical and numerical studies of outflow-dense-medium interactions \citep[e.g.,][]{Mou2022,Lei2024,Matsumoto2024,Zhuang2025,Mou2025bow}. Physically, a shallower inner profile (\(k_{\rm i}\lesssim1.5\)) suppresses kinetic-energy dissipation inside \(R_{\rm B}\), so the early flare is weakened while the outer density jump drives a sharper late rise. After this sharp increase, the strongly decelerated ejecta enter a gradual decline as they expand beyond \(R_{\rm B}\). We note that the three-dimensional cold-gas geometries (e.g., clumpy clouds or a dusty torus) with finite covering fractions would smooth the peak, yielding light-curve shapes more similar to the moderate-density cases shown here.

\subsection{Fast and Narrow Jet Scenario} \label{sec:jet}

In addition to the wide-angle, low-velocity wind models, we also explore the interaction between a relativistic, narrow jet and the CNM. The fiducial jet parameters are \(M_{\rm ej}=3\times10^{-3}\,M_{\odot}\), \(v_{\rm ej}=0.9\,c\) (corresponding to kinetic energy \(E_{\rm k} \sim 4\times10^{51}\,\mathrm{erg}\)), and an opening angle \(\theta_{\rm ej}=10^\circ\). We fix the CNM properties to \(R_{\rm B}=7.5\times10^{16}\,\mathrm{cm}\), \(n_{\rm B}=10^{3}\,\mathrm{cm^{-3}}\), \(k_{\rm i}=2.5\) and \(k_{\rm o}=0\). The dynamical evolution is broadly similar to the wind case (Figure~\ref{fig:jet}), although the narrower opening angle suppresses RT instabilities and limits lateral mixing.

The light curves of the jet-CNM interaction are shown in Figure~\ref{fig:jet_lc}(a). Because the jet propagates much faster than the wind, it reaches and crosses the Bondi radius shortly after producing the first peak at higher frequencies (\(87\,\mathrm{GHz}\)), yielding a secondary rise near \(\sim 50\) days. The jet's lower ejecta mass also leads to earlier deceleration than in the wind case, producing an initial decline followed by a second peak at \(\sim 100-200\) days. Notably, when the jet crosses \(R_{\rm B}\) the spectral peak remains high (\(\nu_{\rm p}\sim 40\,\mathrm{GHz}\)), so lower observed frequencies cannot resolve a distinct double peak and instead display a bumpy rise (a shoulder) and a frequency-dependent second peak. 

We also explore the impact of different CNM radial profiles on jet-CNM interaction, as shown in Figure~\ref{fig:jet_lc}(b), by varying the inner and outer density slopes \(k_{\rm i}\) and \(k_{\rm o}\). Increasing \(k_{\rm i}\) raises the luminosity of the first peak and delays its occurrence, thereby making the two peaks less distinct (or more evident shoulder shape). The \(k_{\rm o}\) further modulates the shape of the light curve beyond the Bondi radius.

\begin{figure*} 
  \centering
  \includegraphics[width=1.0\linewidth]{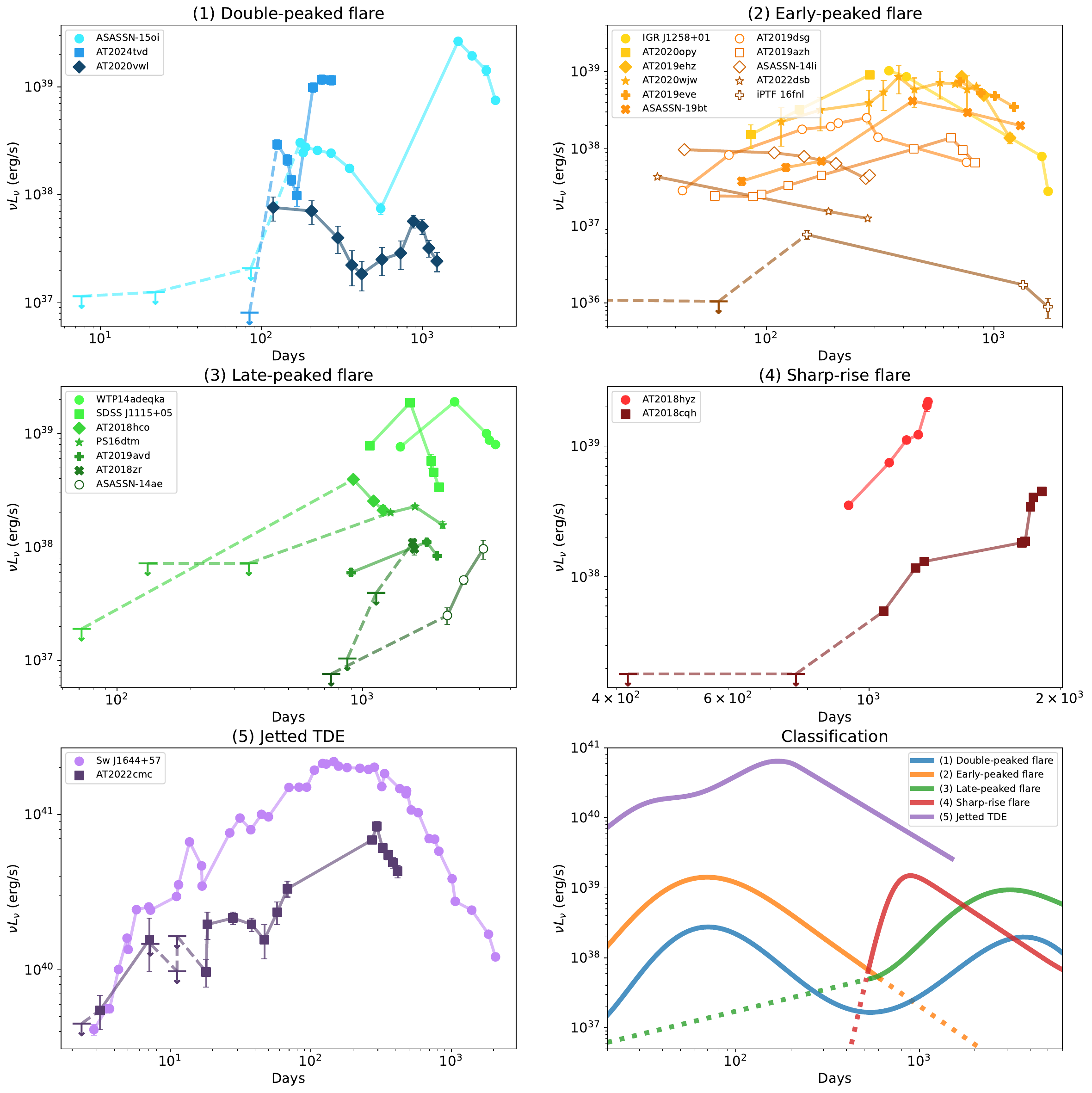}
  \caption{The classification of the typical radio light curves for TDEs, which are tentatively classified as: double-peaked flares (top-left panel), early-peaked flares (\(<\) 2 yr, top-right panel), late-peaked flares (\(>\) 2 yr, middle-left panel), sharp-rise flares (middle-right panel), jetted TDEs (bottom-left panel) and typical features together (bottom-right panel). In the bottom-right panel, the dotted line indicates a possible observational gap, which may arise either from intrinsically weak emission or from incomplete observational coverage. The sources (usually exhibit in 5GHz) and references are (1) The double-peaked sources include ASASSN-15oi \citep{Horesh2021a,Hajela2025-15oi}, AT2024tvd \citep[10\,GHz,][]{Sfaradi2025} and AT2020vwl \citep{Goodwin2023AT2020vwl,Goodwin2024AT2020vwl}. (2) The early-peaked sources include IGR J12580+0134 \citep{Irwin2015,Perlman2017}, AT2020opy \citep{Goodwin2023}, AT2019ehz \citep{Cendes2024}, AT2020wjw/eRASSt J234403-352640 \citep{Goodwin2024-J2344}, AT2019eve \citep{Cendes2024}, ASASSN-19bt \citep{Christy2024ASASSN-19bt}, AT2019dsg \citep{Cendes2021-19dsg,Mohan2022,Cendes2024}, AT2019azh \citep{Goodwin2022-19azh}, ASASSN-14li \citep{Alexander2016}, AT2022dsb \citep{Malyali2024-22dsb}, and iPTF 16fnl \citep{Horesh2021b,Cendes2024}. (3) The late-peaked sources include WTP14adeqka \citep[3\,GHz,][]{Golay2025}, SDSS J1115+0544 \citep{Zhang2025-SDSSJ1115}, AT2018hco \citep{Cendes2024}, PS16dtm \citep{Cendes2024}, AT2018zr \citep{Cendes2024}, AT2019avd \citep{Wang2023-19avd,Goodwin2025-22eROSAT_sample}, and ASASSN-14ae \citep{Cendes2024}. (4) The sharp-rise sources include AT2018hyz \citep{Cendes2022,Sfaradi2024} and AT2018cqh \citep[1\,GHz,][]{Zhang2024-18cqh}. (5) The jetted TDE include Swift J1644+57 \citep{Zauderer2011,Berger2012,Zauderer2013,Eftekhari2018,Cendes2021} and AT2022cmc \citep{Andreoni2022,Pasham2023,Rhodes2023,Rhodes2025}. }
  \label{fig:Classification}
\end{figure*}

\section{Discussion} \label{sec:discussion}

 To qualitatively compare our simulations with observational results, we summarize the main features of TDE radio light curves in Figure~\ref{fig:Classification} and Table~\ref{tab:radio_tde_models}. These features are simply grouped into five categories: double-peaked flares; early-peaked flares (\(<2\,\mathrm{yr}\)); late-peaked flares (\(>2\,\mathrm{yr}\)); sharp-rise flares (e.g., steeper than \(\nu L_{\nu}\propto t^{5}\)); and jetted TDEs. We note that this classification may be incomplete because of limited radio coverage. In the following subsections we describe each category, compare them with our simulation results, and discuss the corresponding physical interpretations.

\subsection{Double-peaked Flare} \label{sec:doublepeak}

Several TDEs show clear double-peaked radio light curves, which include ASASSN-15oi, AT2020vwl and AT2024tvd \citep[e.g.,][]{Goodwin2024AT2020vwl,Hajela2025-15oi,Sfaradi2025}. In our simulations, this shape can be naturally reproduced when the CNM radial profile transitions near the Bondi radius \citep[see also][for analytical work]{Matsumoto2024}: the first peak arises from the outflow-CNM interaction inside \(R_{\rm B}\), while the second peak originates from the interaction beyond \(R_{\rm B}\). Such distinct CNM distributions inside and outside the Bondi radius are expected, being governed primarily by the SMBH potential and the host bulge potential, respectively. Notably, the early peaks of AT2024tvd and ASASSN-15oi exhibit both steep rises and steep decays, implying large inner density slopes \(k_{\rm i}\) in their host galaxies. While the steep slope of AT2024tvd may be related to its special off-nuclear environment, this preference of double-peaked events for extreme inner CNM distributions warrants further investigation. We note that two radio peaks can also result from two separate outflow episodes \citep[e.g.,][]{Goodwin2024AT2020vwl,Hajela2025-15oi,Wu2026,Sato2025}, although the physical origin of multiple ejection episodes remains uncertain. We also perform a tentative fit to the radio light curve and broadband spectrum of AT2020vwl using our simulations, and our model can well reproduce the evolution of both the light curve and spectrum (see Appendix~\ref{sec:fit} for more details).

\subsection{Single-peaked Flare} \label{sec:singlepeak}

A substantial fraction of radio TDEs show a single radio peak in the currently available datasets (see Figure~\ref{fig:Classification}; e.g., \citealt{Alexander2020,Cendes2024}), although sparse late-time coverage (particularly beyond \(\sim10^{3}\) days) leaves the possibility of undetected rebrightenings. The observed single peak is commonly interpreted as synchrotron emission from the interaction between the TDE-driven outflow and a single power-law CNM \citep[e.g.,][]{Alexander2020,Alexander2025,Hu2025ApJ,Mou2025}. Several radio TDEs exhibit peak times shorter than two years, or even within \(\sim10^{2}\) days. Such early peaks naturally arise from the outflow-CNM interaction inside the Bondi radius in our simulations, and the potential second peak may be weak or abscent if the CNM gas is tenuous beyond the Bondi radius (e.g., \(k_{\rm o}\gtrsim1\); see the bottom-left panel of Figure~\ref{fig:kb}). Furthermore, if the deceleration radius is comparable to or smaller than the Bondi radius (\(M_{\rm ej}/M_{\rm B}\lesssim1\)), either because of a high inner CNM density or a low ejecta mass, the second peak should be strongly suppressed, leaving only a single early-peaked flare.

In recent years, several TDEs have exhibited extremely late-time radio flares, emerging two years or more after the initial optical/X-ray discovery \citep[e.g.,][]{Horesh2021a,Horesh2021b,Cendes2022,Goodwin2023,Cendes2024}. These late flares are primarily produced by the interaction of the outflow with the CNM beyond the Bondi radius. A deficit of gas inside the Bondi radius, either because of a lower density or a shallow inner profile (\(k_{\rm i}\lesssim 1.5\)), suppresses the first peak and naturally yields a single late-time peak in our simulations. Such a shallow inner environment, that consistent with X-ray observations of nearby galactic nuclei including M87, NGC~3115, and NGC~1600 \citep{Wong2014,Russell2015,Runge2021}, also tends to produce steeper late-time rises. The particularly steep rises as observed in AT2018zr and ASASSN-14ae may reflect local density bumps near the Bondi radius with \(k_{\rm o}\gtrsim 0.0\). An alternative explanation is a delayed outflow launch during a low-activity state of the SMBH, although the physical origin of such delays remains unclear \citep[e.g.,][]{Horesh2021a,Sfaradi2022,Cendes2022,Cendes2025AT2018hyz,Sato2025}. Notably, very long delays can produce apparently steeper light curves as a consequence of an offset in the assumed time origin \citep[e.g.,][]{Cendes2022,Cendes2025AT2018hyz,Matsumoto2025}. An off-axis jet can also produce a delayed onset, but it typically predicts a much steeper rise \citep[e.g.,][]{Lei2016,Matsumoto2023,Sfaradi2024}. We further discuss this scenario in Section~\ref{sec:sharprise}. Finally, our distinction between ``early'' and ``late'' peaks, based on a timescale of two years, is a crude classification that roughly separates the observed and simulated double-peak timescales \citep[][]{Cendes2024}, and will require more observations to validate.

\subsection{Sharp-rise Flare} \label{sec:sharprise}

AT2018hyz and AT2018cqh were all detected in the radio band more than a thousand days after their TDE discoveries \citep{Cendes2022,Sfaradi2024,Zhang2024-18cqh,Cendes2024}. Unlike the single late-time peaks discussed above, these sources exhibit extremely steep rises, in some cases even steeper than \(\nu L_{\nu}\propto t^{5}\). One explanation invokes interaction between the outflow and a dense gas structure at large radii, such as clumpy cold clouds or a dusty torus \citep[e.g.,][]{Mou2022,Bu2023,Lei2024,Matsumoto2024,Zhuang2025,Mou2025bow,Yang2025}, where the sharp rise reflects a sudden increase in ambient density. This behavior is reproduced in our simulations (see Figure~\ref{fig:kcold}). The sharp-rise events with extremely steep late flares typically lack early radio emission, which may be caused by the preference for a shallow inner profile ($k_{\rm i}\lesssim1.5$). An alternative explanation is an off-axis relativistic jet initially misaligned with our line of sight: relativistic beaming suppresses early emission, producing a sharp radio rise as the jet decelerates and the beaming cone widens into view \citep[e.g.,][]{Lei2016,Matsumoto2023,Sfaradi2024,Sato2024,Lu2024,Matsumoto2025}. Note that the required jet energy and Lorentz factors may be extreme for some events \citep[e.g.,][]{Sfaradi2024,Sato2024}, and the low observed incidence of jetted TDEs disfavors this scenario as a universal explanation for all sharp-rise flares \citep[e.g.,][]{Teboul2023}.

\subsection{Jetted TDE} \label{sec:jettde}
Jetted TDEs, although still limited in number, constitute an important subclass of radio TDEs \citep[e.g.,][]{Zauderer2011,Berger2012,Kelley2014,De_Colle2020,Andreoni2022}. Their radio light curves often exhibit pre-peak bumps or shoulder-like shapes (see the bottom-left panel of Figure~\ref{fig:Classification}; e.g., \citealt{Cendes2021,Rhodes2025}). At sufficiently high frequencies, some jetted events (e.g., Swift J1644+57) show distinct double peaks \citep[e.g.,][]{Eftekhari2018,Cendes2021}. Previous studies have interpreted these features with two-component jet models composed of a narrow, fast cocoon and a wider, slower sheath, whose interaction with a stratified CNM can roughly reproduce the observed jetted-TDE light curves \citep[e.g.,][]{Wang2014,Mimica2015,Liu2015,Teboul2023,Zhou2024,Sato2025}. In our simulations, both the low-frequency shoulder features and the high-frequency double peaks (typically \(\gtrsim 20\,\mathrm{GHz}\)) are consistent with observations. Variations in the CNM radial structure further modulate the light-curve shape \citep[see the right panels of Figure~\ref{fig:jet_lc}; e.g.,][]{De_Colle2012,Generozov2017}, and our models are broadly consistent with the observed behavior of Swift J1644+57 and AT2022cmc. Jet precession or angular structure can further distort the timing and shape of the radio emission \citep[e.g.,][]{Stone2012,Teboul2023,Calderon2024,Lu2024,Wang2025,Yuan2026}. A detailed, quantitative comparison that constrains the CNM will require a broader parameter survey, which we plan to pursue in future work.

\startlongtable
\setlength{\tabcolsep}{0pt} 
\begin{deluxetable*}{llll}
\tabletypesize{\footnotesize}
\tablecaption{Radio light-curve features of TDEs and possible models in the literature \label{tab:radio_tde_models}}
\tablewidth{0pt}
\tablehead{
{Classification} & \makecell[l]{Implication in our simulations} & \makecell[l]{Proposed models} & \makecell[l]{References} 
}
\startdata
Double-peaked flare& \makecell[l]{interaction between outflow \\ and a broken power-law CNM} & \makecell[l]{(i) \,\,structured CNM [A] \\ (ii) \,multiple outflow episodes [A]} & \makecell[l]{ref. 1 \\ ref. 2} \\
\hline
Early-peaked flare& \makecell[l]{deceleration within \(R_{\rm B}\) or \\ shallow CNM profile outside \(R_{\rm B}\)} & \makecell[l]{(i) \,\,interaction between outflow and\\ \ \ \,\,\,\, a single power-law CNM [A+S]} & \makecell[l]{ref. 3 }  \\
\hline
Late-peaked flare& \makecell[l]{shallow CNM profile inside \(R_{\rm B}\)} & \makecell[l]{(i) \,\,delayed outflow [A]\\ (ii)\, off-axis jet [A+S]} & \makecell[l]{ref. 4 \\ ref. 5}  \\
\hline
Sharp-rise flare& \makecell[l]{dense gas structure (e.g., clouds/torus)} & \makecell[l]{(i) \,\,outflow-dense-gas interaction [A+S] \\ (ii) \,long-delayed outflow [A] \\ (iii) off-axis jet [A+S]} & \makecell[l]{ref. 6 \\ ref. 7 \\ ref. 5}  \\
\hline
Jetted TDE & \makecell[l]{interaction between jet and a\\ broken power-law CNM} & \makecell[l]{(i) \,\,structured CNM [A+S]\\ (ii) \,two-component jet [A+S]\\ (iii) jet precession [A+S]} & \makecell[l]{ref. 8 \\ ref. 9 \\ ref. 10}  \\
\enddata
\tablecomments{The classification of TDE radio flares, implications from our simulations, and proposed models in literature (A: analytical work; S: simulation). For convenience, we list representative references for each model, focusing on earlist and influential studies. A concise description and additional references are given in Section~\ref{sec:discussion}. References: (1) \citet{Matsumoto2024}; (2) \citet{Goodwin2024AT2020vwl,Hajela2025-15oi}; (3) \citet{Alexander2016,Krolik2016,Hu2025ApJ,Mou2025}, see \citet{Alexander2020} for a review; (4) \citet{Horesh2021a,Sfaradi2022,Alexander2025}; (5) \citet{Lei2016,Generozov2017,Matsumoto2023,Sato2024}; (6) \citet{Mou2022,Lei2024,Zhuang2025,Mou2025bow}; (7) \citet{Cendes2022,Cendes2025AT2018hyz}; (8) \citet{De_Colle2012,Generozov2017}; (9) \citet{Wang2014,Mimica2015,Liu2015}; (10) \citet{Teboul2023,Calderon2024}.}
\end{deluxetable*}

\section{Summary} \label{sec:summary}

The increasing complexity of TDE radio light curves motivates a detailed modeling beyond analytical approximations. This work presents three-dimensional hydrodynamic simulations of TDE outflow-CNM interactions using a broken power-law density profile, and computes the resulting radio emission, and provides analytical estimates to aid interpretation. We find that different shape of radio light curves reflect different properties of the CNM and outflow. We summarize the main observational features of TDE radio light curves and present corresponding hydrodynamic simulations that capture their key physical features. The principal results are as follows.

\begin{enumerate}
  \item \textbf{Double-peaked flare:} Interaction between an outflow and a broken power-law CNM naturally produces double-peaked radio light curves, with characteristic peaks at \(\sim\) several hundred days and \(\sim\) several thousand days respectively. The simulations are qualitatively consistent with analytical expectations.
  \item \textbf{Early-peaked flare:} The late flare is weak or absent when the outflow is strongly decelerated within the Bondi radius or when the CNM beyond the Bondi radius is tenuous.
  \item \textbf{Late-peaked flare:} A low-density CNM inside the Bondi radius suppresses the early radio flare, while a late radio flare can arise from the subsequent interaction between the outflow and the CNM beyond \(R_{\rm B}\).
  \item \textbf{Sharp-rise flare:} Interaction between the outflow and dense CNM (for example, dense clouds or a dusty torus) produces rapid accumulation of shocked material and a steep rise in radio luminosity; a shallow inner profile (\(k_{\rm i}\lesssim1.5\)) further weakens the early flare and makes an apparently single late sharp rise more likely.
  \item \textbf{Jetted TDE:} A fast, narrow jet interacting with a broken power-law CNM produces a low-frequency shoulder (a bumpy rise) and double peaks at sufficiently high observed frequencies.
\end{enumerate}

Therefore, the radio emission provides a valuable diagnostic of nuclear gas distributions and outflow properties.

\bibliography{cite}{}

\bibliographystyle{aasjournal}

\begin{appendix}
\renewcommand{\thefigure}{A\arabic{figure}}
\setcounter{figure}{0}
\renewcommand{\thetable}{A\arabic{table}}
\setcounter{table}{0}

\section{Numerical Method} \label{app:setup}

\subsection{Hydrodynamical Equations} \label{sec:hydrodynamics}

We employ the publicly available code \texttt{Athena++} \citep[e.g.,][]{Stone2020} to solve the time-dependent equations of hydrodynamics, 
\begin{equation}
\frac{\partial \rho}{\partial t} + \nabla \cdot (\rho \mathbf{v}) = 0,
\label{eq:continuity}
\end{equation}

\begin{equation}
\frac{\partial (\rho \mathbf{v})}{\partial t} 
+ \nabla \cdot \bigl(\rho\,\mathbf{v} \otimes \mathbf{v} + P\,\mathbf{I}\bigr)
= -\,\rho\,\nabla \Phi,
\label{eq:momentum}
\end{equation}

\begin{equation}
\frac{\partial E}{\partial t} 
+ \nabla \cdot \bigl[(E + P)\,\mathbf{v}\bigr]
= -\,\rho\,\mathbf{v}\cdot\nabla\Phi,
\label{eq:energy}
\end{equation}
where $\mathbf{I}$ is the identity tensor. The total gas energy density is given by \(E = P/(\gamma - 1) + \frac{1}{2} \rho \mathbf{v}^2\), and the gas temperature by \(T = P\mu / (R_{\rm ideal} \rho)\), where the adiabatic index is \(\gamma = 5/3\), the mean molecular weight is \(\mu = 0.6\), and \(R_{\rm ideal}\) is the ideal gas constant. The gravitational potential of the central black hole is included as \(\Phi(r) = -GM_{\rm BH}/r\).

\subsection{Synchrotron Radiation} \label{subsec: SynchrotronRadiation}

Outflow-CNM interactions drive shocks that accelerate electrons. A fraction, $\epsilon_{\rm e} m_{\rm p} (p-2) (\Gamma_{\rm s}-1) / \left[ m_{\rm e} (p-1) (\gamma_{\rm m}-1) \right]$, of the shocked electrons is injected into a power-law distribution, $dn_{\rm nt}(\gamma_{\rm e}) = A\,\gamma_{\rm e}^{-p}\,d\gamma_{\rm e}$, with a minimum electron Lorentz factor ($\gamma_{\rm m} = \max\left\{2,\; 1 + \epsilon_{\rm e} m_{\rm p} (p-2) (\Gamma_{\rm s}-1) / \left[ m_{\rm e} (p-1) \right]\right\}$) and a fraction of thermal energy transferred to non-thermal electrons \citep[$\epsilon_{\rm e}$, e.g.,][]{Sari1999,Huang2003,Sironi2013,Zhang2018}. Here \(\Gamma_{\rm s} = 1/\sqrt{1-(v_{\rm s}/c)^2}\) is the bulk Lorentz factor of the shocked material in each cell. In our simulations, all wind models remain in the deep-Newtonian phase with \(\gamma_{\rm m}=2\), while in the jet models \(\gamma_{\rm m} \sim 80\) is reached with \(v_{\rm ej} = 0.9c\). Shocked gas is identified by computing the entropy \(S\) in each cell and selecting those with \(S > 2\,S_{\rm B}\), where \(S_{\rm B}\) is the entropy at the Bondi radius \citep[e.g.,][]{Colgate1990,Muller2017,Lei2024}. Non-thermal electrons are assumed to acquire a fixed fraction of the post-shock internal energy, reducing sensitivity to turbulent fluctuations. We compute the synchrotron light curve throughout the simulation domain (see \citealt{Lei2024} for details). 

The energy of non-thermal electrons with \(\gamma_{\rm e} > \gamma_{\rm m}\) is 
\begin{equation}
    dE_{\rm nt} = dN_{\text{nt}}m_{\text{e}}c^2\gamma_{\text{m}}\frac{p-1}{p-2} \simeq \epsilon_{\text{e}} U_{\text{s}} ,
    \label{eq:E}
\end{equation}
where \(U_{\rm s} \simeq dN_{\rm s}\,k_{\rm B}\,T_{\rm s}\) represents the total internal energy of shocked electrons in each simulation cell, \(dN_{\rm nt}\) and \(dN_{\rm s}\) denote the total number of non-thermal and shocked electrons respectively, \(k_{\rm B}\) is the Boltzmann constant, \(T_{\rm s}\) is the post-shock temperature. The shocks will also amplify the magnetic fields in the shocked material \citep[e.g.,][]{Reynolds1981}. The magnetic field of shocked material is estimated with
\begin{equation}
    \frac{B^2}{8\pi} = \epsilon_\text{B} u_{\rm s} = \epsilon_\text{B} n_{\text{s}} k_{\text{B}} T_{\rm s},
    \label{eq:B}
\end{equation}
where $n_{\rm s}$ is the number density of shocked materials and $\epsilon_\text{B}$ is the fraction of thermal energy that goes to magnetic field.

In the spectral calculation, a line of sight perpendicular to the equatorial plane is assumed. We simulate only the hemisphere (\(\theta<\pi/2\)) and, by symmetry of the outflow, mirror the simulation results to \(\theta\in[0,\pi]\). The synchrotron self-absorption (SSA) frequency, \(\nu_{\rm a}\), is defined by the optical depth condition \(\int_l \alpha_\nu \, dl = 1\), where \(\alpha_\nu\) is the synchrotron absorption coefficient integrated over all shocked cells along the line-of-sight path through the simulation domain \citep[e.g.,][]{Rybicki1979}. In our parameter space the spectral ordering \(\nu_{\rm m} < \nu_{\rm a}\) holds in almost all cells, so the synchrotron spectrum in most cells can be described by
\begin{equation}
    dL_\nu = 
    \begin{cases}
        dL_{\nu_{\text{a}}} \left( \frac{\nu_{\text{m}}}{\nu_{\text{a}}} \right)^{\frac{1}{2}} \left( \frac{\nu}{\nu_{\text{a}}} \right)^2 & , \quad \nu \leq \nu_{\text{m}} ; \\
        dL_{\nu_{\text{a}}} \left( \frac{\nu}{\nu_{\text{a}}} \right)^{\frac{5}{2}} & , \quad \nu_{\text{m}} \leq \nu \le \nu_{\text{a}} ;  \\
        dL_{\nu_{\text{a}}} \left( \frac{\nu}{\nu_{\text{a}}} \right)^{\frac{1-p}{2}} & , \quad \nu_{\text{a}} \leq \nu , 
    \end{cases}
    \label{eq:L_nu}
\end{equation}
where $dL_{\nu_{\text{a}}} = dL_{\nu_{\text{m}}} (\nu_{\text{a}}/\nu_{\text{m}})^{(1-p)/2}$ with \(dL_{\nu_{\text{m}}} \simeq dN_{\text{nt}} (4/3)\sigma_Tc\gamma_{\text{m}}^2(B^2/8\pi)/{\nu_{\text{m}}}\) and 
\( \nu_{\text{m}} = eB\gamma_{\text{m}}^2/(2\pi m_{\text{e}}c) \). Cells with \(\nu_{\rm a} < \nu_{\rm m}\) are also accounted for using the appropriate synchrotron spectral formula, but they contribute negligibly to the total luminosity. The total synchrotron luminosity of shocked material is then \(L_{\nu,\mathrm{total}}=\sum dL_\nu\). For the relativistic jet models, a Doppler factor \(\delta\) is applied to transform the emission and absorption coefficients into the observer frame, i.e., \(\nu = \delta\nu'\), \(L_\nu = \delta^3 L'_{\nu'}\).

\subsection{Analytical Estimations of Radio Light Curve Features} \label{sec:analytic}

In this section we provide analytical estimates for key timescales and luminosity scalings of the radio light curves, following the spherically symmetric thin-shell framework of \citet{Matsumoto2024}. Within this framework, the synchrotron luminosity in the optically thick/thin regimes and the SSA frequency scale as
\begin{equation}
    \begin{aligned}
      (\nu L_\nu)_{\rm thick} \propto & \,\, \epsilon_{\rm B}^{-1/4}\, n^{-1/4}\, R^2\, T_{\rm s}^{-1/4}\, \nu^{7/2}\, \Omega_{\rm ej}\,  ,\\
      (\nu L_\nu)_{\rm thin} \propto & \,\, \bar\epsilon_{\rm e}\, \epsilon_{\rm B}^{(p+1)/4}\, n_{\rm s}^{(p+5)/4}\, R^3\, v_{\rm s}^{2}\, T_{\rm s}^{(p+1)/4}\, \nu^{(3-p)/2}\, \Omega_{\rm ej}\,  ,\\            
      \nu_{\rm a} \propto & \,\, \bar\epsilon_{\rm e}^{2/(p+4)}\, \epsilon_{\rm B}^{(p+2)/[2(p+4)]}\, n_{\rm s}^{(p+6)/[2(p+4)]}\, v_{\rm s}^{4/(p+4)}\, T_{\rm s}^{(p+2)/[2(p+4)]}\, R^{2/(p+4)}\, ,
    \end{aligned}
\end{equation}
where \(\bar\epsilon_{\rm e} = \epsilon_{\rm e} (p-2)/(p-1)\). The post-shock temperature scales as \(T_{\rm s} \propto v_{\rm s}^2 \propto v_{\rm ej}^2\), a relation that holds in our simulations until the RT instability becomes significant. For an inner power-law CNM density profile \(n = n_{\rm B} (R/R_{\rm B})^{-k_{\rm i}}\) and a freely expanding outflow (\(R\propto t\)), the temporal evolution in these regimes is
\begin{equation}
  \begin{aligned}
    (\nu L_{\nu})_{\rm thick} &\propto t^{(k_{\rm i}+8)/4}, \\
    (\nu L_{\nu})_{\rm thin} &\propto t^{[12-k_{\rm i}(p+5)]/4}, \\
    \nu_{\rm a} &\propto t^{[4-k_{\rm i}(p+6)]/[2(p+4)]} ,    
    \label{eq:slope}
  \end{aligned}
\end{equation}
For our fiducial inner slope \(k_{\rm i}=2.5\) and \(p=2.5\), we obtain \((\nu L_{\nu})_{\rm thick} \propto t^{2.63}\), \((\nu L_{\nu})_{\rm thin} \propto t^{-1.69}\), and \(\nu_{\rm a} \propto t^{-1.33}\). These temporal slopes agree with the early evolution of our simulated light curves (see the left panels of Figure~\ref{fig:w120}), where the actual peak frequency \(\nu_{\rm p}\) is taken as the emission-weighted average of the SSA frequencies over all shocked cells.

The first radio peak typically occurs within the Bondi radius when the synchrotron spectrum transitions from optically thick to thin. The shock radius \(R_{\rm 1st}\) at which this transition occurs is obtained by equating the SSA frequency to the observed frequency, \(\nu = \nu_{\rm a}\), yielding
\begin{equation}
  R_{\rm 1st} \propto \bar\epsilon_{\rm e}^{4/D_{\rm i}}\, \epsilon_{\rm B}^{(p+2)/D_{\rm i}}\, n_{\rm B}^{(p+6)/D_{\rm i}}\, R_{\rm B}^{k_{\rm i}(p+6)/D_{\rm i}}\, v_{\rm ej}^{2(p+6)/D_{\rm i}}\,\nu^{-2(p+4)/D_{\rm i}} ,
\end{equation}
where \(D_{\rm i} \equiv k_i(p+6)-4\). The corresponding peak time and luminosity scale as
\begin{equation}
  \begin{aligned}
    t_{\rm 1st} = \frac{R_{\rm 1st}}{v_{\rm ej}} \propto & \,\, \bar\epsilon_{\rm e}^{4/D_{\rm i}}\, \epsilon_{\rm B}^{(p+2)/D_{\rm i}}\, n_{\rm B}^{(p+6)/D_{\rm i}}\, R_{\rm B}^{k_{\rm i}(p+6)/D_{\rm i}}\, v_{\rm ej}^{[2(p+6)-k_{\rm i}(p+6)]/D_{\rm i}}\, \nu^{-2(p+4)/D_{\rm i}} ,\\
    (\nu L_\nu)_{\rm 1st} \propto & \,\, \Omega_{\rm ej}\, \bar\epsilon_{\rm e}^{(8+k_{\rm i})/D_{\rm i}}\, \epsilon_{\rm B}^{(2p+5-k_{\rm i})/D_{\rm i}}\, n_{\rm B}^{(2p+13)/D_{\rm i}}\, R_{\rm B}^{k_{\rm i}(2p+13)/D_{\rm i}}\, v_{\rm ej}^{2(2p+13)/D_{\rm i}}\, \nu^{[k_{\rm i}(3p+19)-4(p+6)]/D_{\rm i}} .
  \end{aligned}
\end{equation}
For \(k_{\rm i}=2.5\) and \(p=2.5\), these scalings become \(t_{\rm 1st} \propto \bar\epsilon_{\rm e}^{0.23}\, \epsilon_{\rm B}^{0.26}\, n_{\rm B}^{0.49}\, R_{\rm B}^{1.23}\, v_{\rm ej}^{-0.25}\, \nu^{-0.75}\) and \((\nu L_\nu)_{\rm 1st} \propto \Omega_{\rm ej}\, \bar\epsilon_{\rm e}^{0.61}\, \epsilon_{\rm B}^{0.43}\, n_{\rm B}^{1.04}\, R_{\rm B}^{2.61}\, v_{\rm ej}^{2.09}\, \nu^{1.87}\), generally consistent with our simulated results (see the right panels of Figure~\ref{fig:w120}). Thus higher \(\nu\) yields an earlier and brighter first peak, although the complex structure of the shocked region can cause deviations from the simple analytical scalings. Higher \(n_{\rm B}\) gives a brighter but later first peak, while higher \(v_{\rm ej}\) gives a brighter but slightly earlier peak. The ejecta mass \(M_{\rm ej}\) has only a limited effect on the first peak in our simulations. Based on our fiducial simulation (\(t_{\rm 1st} \sim 97.9\) days, \((\nu L_\nu)_{\rm 1st} \sim 2.2 \times 10^{38}\) erg s$^{-1}$ at 6 GHz), we express the peak time and luminosity in terms of the fiducial values as
\begin{equation}
  \begin{aligned}
    t_{\rm 1st} \simeq & \,\, 92.1 \, \text{days} \, \left( \frac{\bar\epsilon_{\rm e}}{0.133} \right)^{4/D_{\rm i}} \left( \frac{\epsilon_{\rm B}}{0.02} \right)^{(p+2)/D_{\rm i}} \left( \frac{n_{\rm B}}{200\,\text{cm}^{-3}} \right)^{(p+6)/D_{\rm i}} \\
    & \times \left( \frac{R_{\rm B}}{1.5\times10^{17}\,\text{cm}} \right)^{k_{\rm i}(p+6)/D_{\rm i}} \left( \frac{v_{\rm ej}}{0.1c} \right)^{[2(p+6)-k_{\rm i}(p+6)]/D_{\rm i}} \left( \frac{\nu}{6\,\text{GHz}} \right)^{-2(p+4)/D_{\rm i}} ,\\
    (\nu L_\nu)_{\rm 1st} \simeq & \,\, 1.7\times10^{38} \, \text{erg s}^{-1} \, \left( \frac{\Omega}{2\pi} \right)\, \left( \frac{\bar\epsilon_{\rm e}}{0.133} \right)^{(8+k_{\rm i})/D_{\rm i}} \left( \frac{\epsilon_{\rm B}}{0.02} \right)^{(2p+5-k_{\rm i})/D_{\rm i}} \left( \frac{n_{\rm B}}{200\,\text{cm}^{-3}} \right)^{(2p+13)/D_{\rm i}} \\ 
    &\times\left( \frac{R_{\rm B}}{1.5\times10^{17}\,\text{cm}} \right)^{k_{\rm i}(2p+13)/D_{\rm i}} \left( \frac{v_{\rm ej}}{0.1c} \right)^{2(2p+13)/D_{\rm i}} \left( \frac{\nu}{6\,\text{GHz}} \right)^{[k_{\rm i}(3p+19)-4(p+6)]/D_{\rm i}} .
  \end{aligned}
\end{equation}
These expressions approximately describe the first peak times and luminosities in our models.
 
The light-curve break time \(t_{\rm break}\) marks the transition from the early decline to the late rise and occurs approximately when the outflow crosses the Bondi radius. For a freely expanding outflow, \(t_{\rm break} \simeq R_{\rm B}/v_{\rm ej}\), but deceleration by the swept-up CNM mass modifies this timescale. Assuming energy conservation, the outflow velocity at radius \(R\) is
\begin{equation}
  v(R) = v_{\rm ej}\left[\frac{M_{\rm ej}}{M_{\rm ej}+M_{\rm sw}(R)}\right]^{1/2},
\end{equation}
where \(M_{\rm sw}(R)\) is the swept-up mass within \(R\). For the broken power-law CNM profile (Equation~\eqref{eq:nR}) with \(k_{\rm i} < 3.0\) and \(k_{\rm o} < 3.0\),
\begin{equation}
  M_{\rm sw}(R)=
  \begin{cases}
    M_{\rm B} \left(\dfrac{R}{R_{\rm B}}\right)^{3-k_{\rm i}}, & R\le R_{\rm B},\\[10pt]
    M_{\rm B} \left[ 1 + \dfrac{3-k_{\rm i}}{3-k_{\rm o}} \left( \left(\dfrac{R}{R_{\rm B}}\right)^{3-k_{\rm o}} - 1 \right) \right], & R>R_{\rm B},
  \end{cases}
\end{equation}
where \(M_{\rm B}=M_{\rm sw}(R_{\rm B})=f_{\rm V} {n_{\rm B} m_{\rm p} \Omega_{\rm ej} R_{\rm B}^{3}}/({3-k_{\rm i}})\) is the swept-up mass at the Bondi radius, and \(f_{\rm V} \equiv (1-\cos^{3}\theta)/(1-\cos\theta)\) is a volume correction factor for lateral sweeping. If \(M_{\rm ej} > M_{\rm B}\), the outflow reaches the break time near the Bondi radius; otherwise, deceleration within \(R_{\rm B}\) suppresses the second flare. For \(M_{\rm ej} > M_{\rm B}\),
\begin{equation}
  t_{\rm break} = \int_0^{R_{\rm B}} \frac{dR}{v(R)}
  = \frac{1}{v_{\rm ej}}\int_0^{R_{\rm B}} \left[1 + \frac{M_{\rm sw}(R)}{M_{\rm ej}}\right]^{1/2} dR
  = \frac{R_{\rm B}}{v_{\rm ej}} \int_0^1 \left[1 + \frac{M_{\rm B}}{M_{\rm ej}} x^{3-k_{\rm i}}\right]^{1/2} dx ,
\end{equation}
where \(x \equiv R / R_{\rm B}\) is the dimensionless radius normalized by the Bondi radius. This integral can be expressed in terms of the Gauss hypergeometric function as
\begin{equation}
  t_{\rm break} = \frac{R_{\rm B}}{v_{\rm ej}}\,
  {}_2F_1\!\left(-\frac{1}{2},\frac{1}{3-k_{\rm i}};1+\frac{1}{3-k_{\rm i}};-\frac{M_{\rm B}}{M_{\rm ej}}\right).
\end{equation}
The break-time luminosity is approximated by \((\nu L_\nu)_{\rm break} \simeq (\nu L_\nu)_{\rm 1st}(t_{\rm break}/t_{\rm 1st})^{[12-k_{\rm i}(p+5)]/4}\). Using our typical simulation parameters, we obtain
\begin{equation}
  \begin{aligned}
  t_{\rm break} \simeq & \,\, 579.1 \, \text{days} \, \left( \frac{v_{\rm ej}}{0.1c} \right)^{-1} \left( \frac{R_{\rm B}}{1.5 \times 10^{17} \, \text{cm}} \right) \\
  & \times \left[ 1 + \frac{0.156}{(3 - k_{\rm i})(4 - k_{\rm i})} \left( \frac{f_{\rm V}}{1.75} \right) \left( \frac{\Omega}{2\pi} \right)  \left( \frac{n_{\rm B}}{200 \, \text{cm}^{-3}} \right) \left( \frac{R_{\rm B}}{1.5 \times 10^{17} \, \text{cm}} \right)^3 \left( \frac{M_{\rm ej}}{0.02 \, M_\odot} \right)^{-1} \right], \\
  (\nu L_\nu)_{\rm break} \simeq & \,\, L_{\rm break,fid}^{(E)}(k_{\rm i},p) \, \left( \frac{\Omega}{2\pi} \right)\left( \frac{\bar\epsilon_{\rm e}}{0.133} \right)\left( \frac{\epsilon_{\rm B}}{0.02} \right)^{\frac{p+1}{4}} \\
  & \times \left( \frac{n_{\rm B}}{200\,\text{cm}^{-3}} \right)^{\frac{p+5}{4}}\left( \frac{R_{\rm B}}{1.5\times10^{17}\,\text{cm}} \right)^{3}\left( \frac{v_{\rm ej}}{0.1c} \right)^{\eta_v}\left( \frac{\nu}{6\,\text{GHz}} \right)^{\eta_\nu} \\
  & \times \left[1 + \frac{0.156}{(3 - k_{\rm i})(4 - k_{\rm i})} \left( \frac{f_{\rm V}}{1.75} \right) \left( \frac{\Omega}{2\pi} \right)\left( \frac{n_{\rm B}}{200\,\text{cm}^{-3}} \right)\left( \frac{R_{\rm B}}{1.5\times10^{17}\,\text{cm}} \right)^3\left( \frac{M_{\rm ej}}{0.02\,M_\odot} \right)^{-1}\right]^{\frac{12-k_{\rm i}(p+5)}{4}},
  \end{aligned}
\end{equation}
where \(L_{\rm break,fid}^{(E)}(k_{\rm i},p) = 1.7\times10^{38}\,\text{erg s}^{-1} \left({579.1}/{92.1}\right)^{{[12-k_{\rm i}(p+5)]}/{4}}\), \(\eta_v = (k_{\rm i}p^2+9k_{\rm i}p+20k_{\rm i}-4p+4)/\{2\,[k_{\rm i}(p+6)-4]\}\), and \(\eta_\nu = (-k_{\rm i}p^2-3k_{\rm i}p+18k_{\rm i}+4p)/\{2\,[k_{\rm i}(p+6)-4]\}\). For our fiducial values \(k_{\rm i}=2.5\) and \(p=2.5\), the break luminosity simplifies to
\begin{equation}
  \begin{aligned}
  (\nu L_\nu)_{\rm break} \simeq & \,\, 7.60\times10^{36}\ {\rm erg\ s^{-1}} \, \left( \frac{\Omega}{2\pi} \right)\left( \frac{\bar\epsilon_{\rm e}}{0.133} \right)\left( \frac{\epsilon_{\rm B}}{0.02} \right)^{0.88} \\
  & \times \left( \frac{n_{\rm B}}{200\,\text{cm}^{-3}} \right)^{1.88}\left( \frac{R_{\rm B}}{1.5\times10^{17}\,\text{cm}} \right)^{3}\left( \frac{v_{\rm ej}}{0.1c} \right)^{3.36}\left( \frac{\nu}{6\,\text{GHz}} \right)^{0.6} \\
  & \times \left[1 + 0.208\left( \frac{f_{\rm V}}{1.75} \right)\left( \frac{n_{\rm B}}{200\,\text{cm}^{-3}} \right)\left( \frac{\Omega}{2\pi} \right)\left( \frac{R_{\rm B}}{1.5\times10^{17}\,\text{cm}} \right)^3\left( \frac{M_{\rm ej}}{0.02\,M_\odot} \right)^{-1}\right]^{-1.69}.
  \end{aligned}
\end{equation}
These expressions capture the dependence of \(t_{\rm break}\) and \((\nu L_\nu)_{\rm break}\) on the key physical parameters and are broadly consistent with our simulations, although the simulated frequency dependence is weaker. The break time decreases with \(v_{\rm ej}\) and \(M_{\rm ej}\), while the break luminosity shows the opposite trend. Both quantities increase with \(n_{\rm B}\). For fixed \(M_{\rm ej}\), a steeper inner profile (larger \(k_{\rm i}\)) gives a later break time because the outflow sweeps up more mass before reaching the Bondi radius.

When the outflow expands beyond the Bondi radius, the radio luminosity rises again due to the change in the CNM density profile. In the optically thin regime, Equation~\eqref{eq:slope} predicts \((\nu L_\nu)_{\rm thin} \propto t^{3}\) and \(\nu_{\rm a} \propto t^{0.3}\), which explains the bumpy evolution of \(\nu_{\rm p}\) seen in our simulations. However, strong deceleration and the finite thickness of the shocked region smooth the transition, resulting in a shallower rise (\(\nu L_\nu \propto t^{1.7}\)) that deviates from the ideal \(t^{3}\) scaling.

The second radio peak time \(t_{\rm 2nd}\) does not coincide exactly with the classical thin-shell deceleration condition \(M_{\rm sw}\sim M_{\rm ej}\). Due to complex hydrodynamics and geometric effects in our simulations, the second peak occurs when only a fraction of the ejecta has entered the effective deceleration phase. We therefore introduce an effective deceleration fraction \(f_{\rm dec}\) via \(M_{\rm sw}(R_{\rm dec}^{\rm eff}) = f_{\rm dec} M_{\rm ej}\), where \(R_{\rm dec}^{\rm eff}\) is the effective deceleration radius. For \(R_{\rm dec}^{\rm eff}>R_{\rm B}\),
\begin{equation}
  R_{\rm dec}^{\rm eff}=R_{\rm B}\left[1+\frac{3-k_{\rm o}}{3-k_{\rm i}}\left(f_{\rm dec}\frac{M_{\rm ej}}{M_{\rm B}}-1\right)\right]^{1/(3-k_{\rm o})}.
\end{equation}
The second peak time is then
\begin{equation}
  \begin{aligned}
    t_{\rm 2nd}=&\int_0^{R_{\rm dec}^{\rm eff}}\frac{dR}{v(R)}=\frac{1}{v_{\rm ej}}\int_0^{R_{\rm dec}^{\rm eff}}\left[1+\frac{M_{\rm sw}(R)}{M_{\rm ej}}\right]dR \\
    =&\frac{R_{\rm B}}{v_{\rm ej}}\Bigg[\frac{R_{\rm dec}^{\rm eff}}{R_{\rm B}}+\frac{M_{\rm B}}{M_{\rm ej}(4-k_{\rm i})} +\frac{M_{\rm B}}{M_{\rm ej}}\left(\frac{k_{\rm i}-k_{\rm o}}{3-k_{\rm o}}\left(\frac{R_{\rm dec}^{\rm eff}}{R_{\rm B}}-1\right)
    +\frac{3-k_{\rm i}}{(3-k_{\rm o})(4-k_{\rm o})}\left[\left(\frac{R_{\rm dec}^{\rm eff}}{R_{\rm B}}\right)^{4-k_{\rm o}}-1\right]\right)\Bigg].
  \end{aligned}
\end{equation}
Our simulations are well described by the empirical relation \(f_{\rm dec} \simeq 0.5455+0.4897\,M_{\rm B}/M_{\rm ej}\), which ranges from 0.5 to 1.0. For the fiducial parameters in Section~\ref{sec:dynamic}, this gives \(f_{\rm dec}\simeq 0.82\). For \(k_{\rm i}=2.5\) and \(k_{\rm o}=0\), the expression simplifies to
\begin{equation}
  t_{\rm 2nd} = \frac{R_{\rm B}}{v_{\rm ej}}\Bigl[F_{\rm D}+\frac{M_{\rm B}}{M_{\rm ej}}\Bigl(\frac{5}{6}F_{\rm D}+\frac{1}{24}F_{\rm D}^4-\frac{5}{24}\Bigr)\Bigr],
\end{equation}
where \(F_{\rm D} = \left(6f_{\rm dec}{M_{\rm ej}}/{M_{\rm B}}-5\right)^{1/3}\). With our fiducial parameters, we obtain \(t_{\rm 2nd} \sim 1710\) days, consistent with the simulation. In general, larger \(M_{\rm ej}\) or smaller \(M_{\rm B}\) delays the second peak, while larger \(v_{\rm ej}\) advances it. The corresponding luminosity can be estimated as \((\nu L_\nu)_{\rm 2nd} \simeq (\nu L_\nu)_{\rm break}(t_{\rm 2nd}/t_{\rm break})^{s_{\rm o}}\), but the actual slope \(s_{\rm o}\) is much shallower than the ideal optically thin slope from Equation~\eqref{eq:slope} due to continuous deceleration and must be determined from simulations. Hence we do not provide a detailed analytical expression here. Qualitatively, larger \(n_{\rm B}\), \(v_{\rm ej}\), or \(M_{\rm ej}\) all enhance \((\nu L_\nu)_{\rm break}\) and thus boost the second peak, although \(n_{\rm B}\) and \(v_{\rm ej}\) also shift \(t_{\rm 2nd}\) to earlier times, weakening the net dependence. A smaller \(k_{\rm o}\) increases \(s_{\rm o}\) and leads to a higher luminosity because it reduces \(R_{\rm dec}^{\rm eff}\) and raises the post-shock density.

After the outflow sweeps up enough mass, the radio luminosity decays again due to strong deceleration. In the outer CNM profile, \(R \propto t^{2/(5-k_{\rm o})}\) and \(T_{\rm s} \propto t^{2(k_{\rm o}-3)/(5-k_{\rm o})}\), yielding \(\nu_{\rm a} \propto t^{-(3p+14)/[(5-k_{\rm o})(p+4)]}\) and \((\nu L_\nu)_{\rm thin} \propto t^{3(p+1)/[2(5-k_{\rm o})]}\) for the SSA frequency and luminosity, respectively. For our fiducial parameters \(k_{\rm o}=0\) and \(p=2.5\), these scalings give \(\nu_{\rm a} \propto t^{-0.66}\) and \((\nu L_\nu)_{\rm thin} \propto t^{-1.05}\). However, the RT instability significantly enhances kinetic energy dissipation within the shocked region, leading to a weighted-average shock velocity \(\bar{v}_{\rm s} \propto t^{-1}\) in our simulations. This produces steeper declines, with \(\nu_{\rm a} \propto t^{-0.91}\) and \((\nu L_\nu)_{\rm thin} \propto t^{-1.85}\), which are closer to the trends seen in our simulated light curves.

\section{Application to AT2022vwl} \label{sec:fit}

\begin{figure*} 
  \centering
  \includegraphics[width=1.0\linewidth]{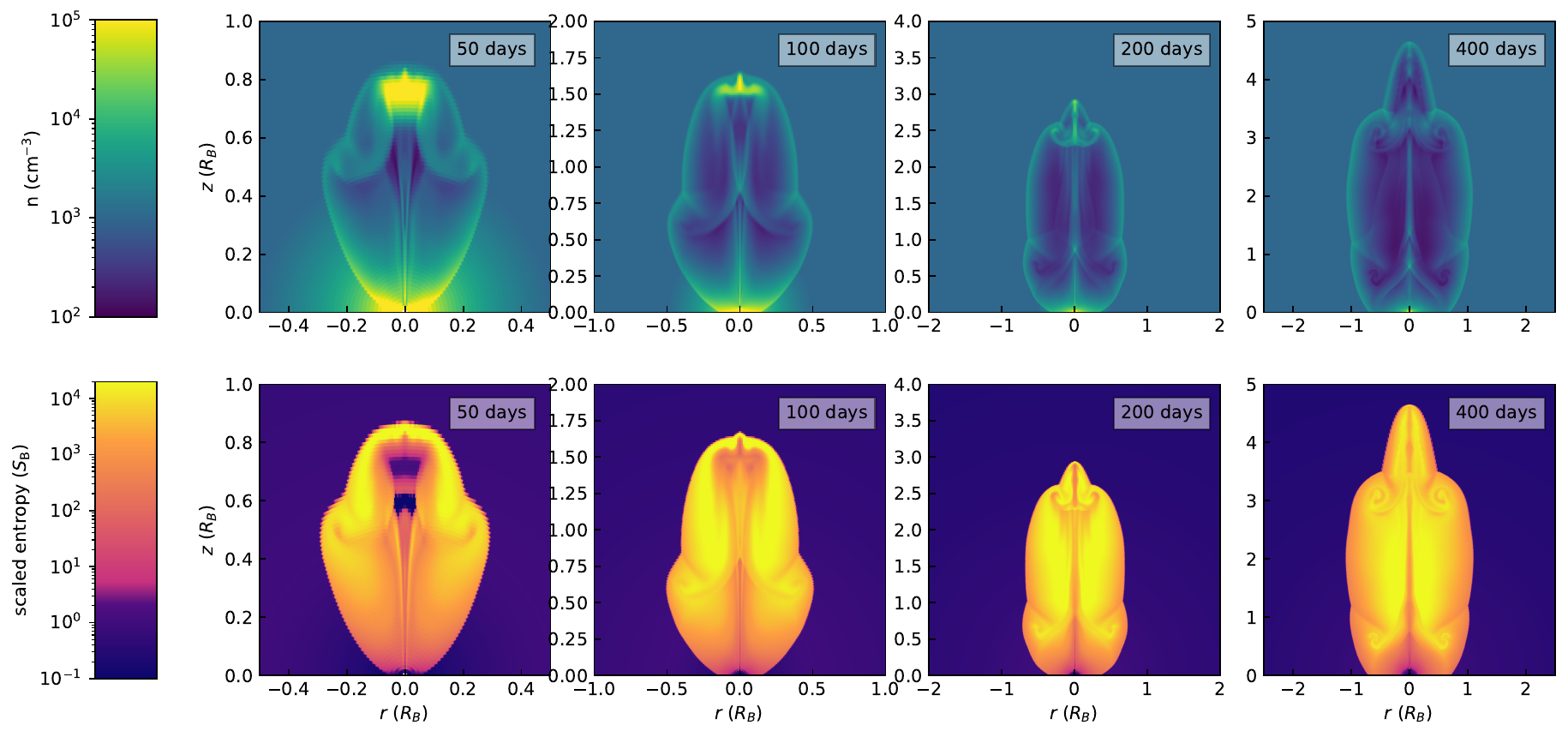}
  \caption{Same as Figure~\ref{fig:w120}, but for the jet scenario.}
  \label{fig:jet}
\end{figure*}

\begin{figure*} 
  \centering
  \includegraphics[width=1.0\linewidth]{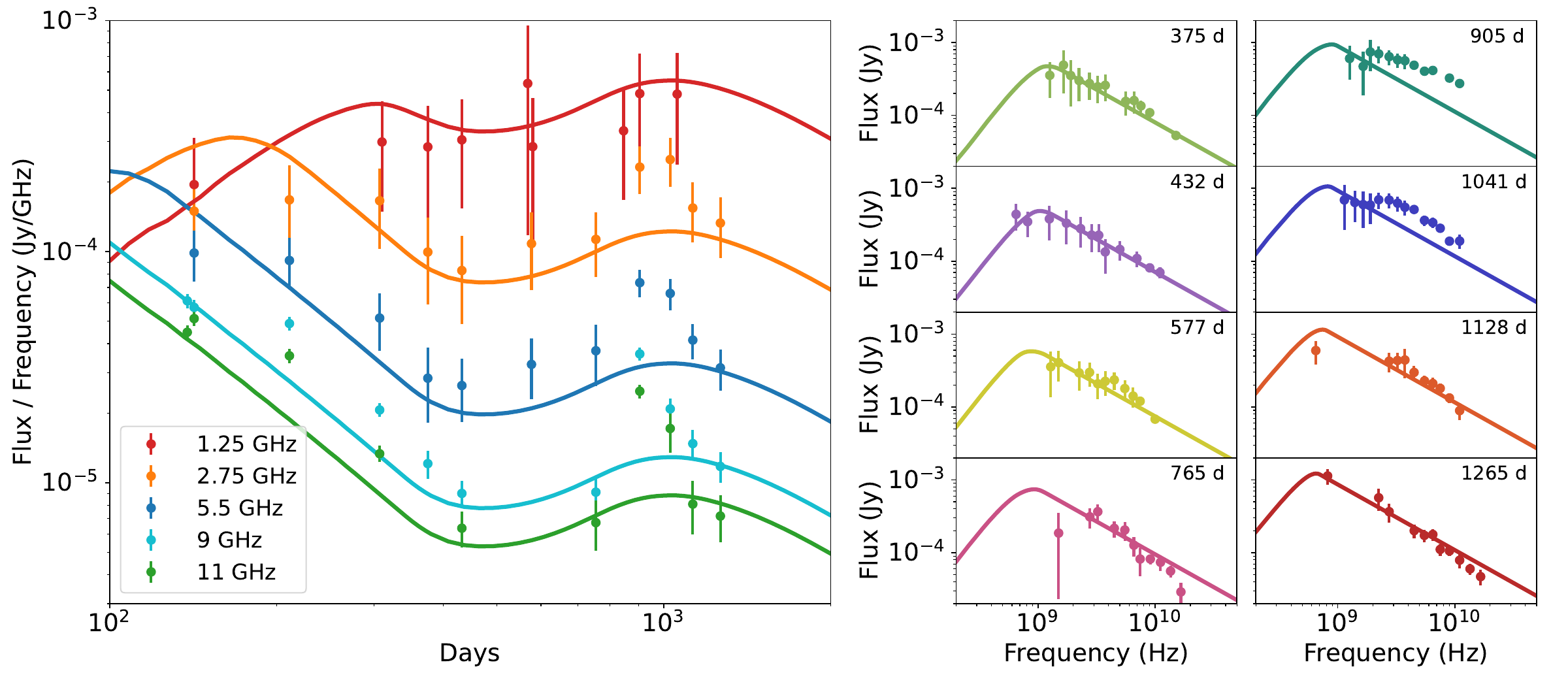}
  \caption{The left panel shows the radio light curves of AT2020vwl at different frequencies of 1.25 GHz (red), 2.75 GHz (orange), 5.5 GHz (blue), 9 GHz (cyan), and 11 GHz (green) (the data is selected from ref. \citep[e.g.,][]{Cendes2024}), where the solid lines are our model predictions. The right panel presents the SED fitting for AT2020vwl at several epochs during the second radio flare.}
  \label{fig:fit}
\end{figure*}

We apply our model to a TDE candidate of AT2020vwl, which is found in a galaxy SDSS J153037.80+265856.8/LEDA 1794348 at z = 0.0325 on October 10, 2020 \citep[e.g.,][]{Hodgkin2020,Hammerstein2021,Yao2023AT2020vwl}. Radio observations started at around 140 days after the optical flare, where the radio was detected in the declining phase \citep[e.g.,][]{Goodwin2023AT2020vwl}. At \(\nu>2\) GHz, the luminosity falls until \(\sim\!400\) days, and the light curve can be explained by the interaction between the wind and the CNM. After 400 days, the light curve enters a second flare lasting 1265 days \citep[e.g.,][]{Goodwin2024AT2020vwl}, which may arise from the density profile transition at the Bondi radius.

To reproduce the light curve of AT2020vwl, we perform hydrodynamical simulations of the outflow-CNM interaction with the wind structure (\(\theta_{\rm ej} = 120\degree\)), motivated by its second radio flare which is notably weaker than the first. The Bondi radius of this source is \(R_{\rm B}  \sim 2.3 \times 10^{16} {\rm cm}\) \citep[e.g.,][]{Bondi1952,Matsumoto2024}, with \(M_{\rm BH}\sim 10^{5.6} M_\odot\) \citep[e.g.,][]{Yao2023AT2020vwl}. Based on the observed spectral energy distribution and light-curve slope, we approximately estimate the $k_{\rm i} = 2.5$ and $p=2.8$ \citep[e.g.,][]{Goodwin2023AT2020vwl,Goodwin2024AT2020vwl}. For simplicity, we assume a flat CNM profile outside the Bondi radius, i.e., $k_{\rm o} = 0.0$, and set \(n_{\rm B} = 2000 ~{\rm cm}^{-3}\) \citep[e.g.,][]{Goodwin2024AT2020vwl}. We find that the total swept-up mass within the Bondi radius is \(M_{\rm B} \sim 2.6 \times 10^{-4}~M_\odot\). Observationally, a distinct double-flare feature requires the total outflow mass to satisfy \(M_{\rm ej} \gtrsim\) a few times \(M_{\rm B}\). Based on observational data, the second radio flare is expected to begin between 432 and 577 days, so we can estimate the outflow velocity \(v_{\text{ej}} \gtrsim 0.022c \). We present the fitting of the radio light curves for AT2022vwl in the left panel of Figure \ref{fig:fit}, where we choose a set of appropriate parameters: \(M_{\rm ej} \sim 7.4 \times 10^{-4} M_\odot\), $v_{\text{ej}} \sim 0.025c$, $\epsilon_{\text{e}} = 0.06$ and $\epsilon_\text{B} = 0.05$. In the right panel of Figure~\ref{fig:fit}, we present the spectral fitting at different epochs. Our model successfully reproduces the slow rise at late times (from $\sim400$ days), which is not well captured by alternative models \citep[e.g.,][]{Zhuang2025,Matsumoto2024}, with only slight deviations observed in two cases between 800 and 1000 days. This steep variation over a period of approximately 100 days may result from the interaction of the outflow with another steep density profile, such as a torus or clouds \citep[e.g.,][]{Zhuang2025,Lei2024}.

\end{appendix}

\end{document}